\documentclass[lettersize,journal]{IEEEtran}

\usepackage{tikz}
\newcommand*\emptycirc[1][0.85ex]{\tikz\draw (0,0) circle (#1);} 
\newcommand*\halfcirc[1][0.85ex]{%
	\begin{tikzpicture}
	\draw[fill] (0,0)-- (90:#1) arc (90:270:#1) -- cycle ;
	\draw (0,0) circle (#1);
	\end{tikzpicture}}
\newcommand*\fullcirc[1][0.85ex]{\tikz\fill (0,0) circle (#1);}

\usepackage{cite}
\usepackage[numbers,sort&compress]{}
\usepackage{amsmath,amssymb,amsfonts}
\usepackage{array}
\usepackage{algorithmic}
\usepackage{graphicx}
\usepackage{makecell}

\usepackage{textcomp}
\usepackage{xcolor}
\usepackage{booktabs}
\usepackage{multirow}
\usepackage{subfigure}
\usepackage{longtable}
\usepackage{threeparttable}
\usepackage{enumitem}
\usepackage{fontawesome}
\usepackage{amssymb}
\usepackage{soul}
\usepackage{amsmath,amssymb,amsfonts}
\usepackage{algorithmic}
\usepackage{graphicx}
\usepackage{textcomp}
\usepackage{xcolor}
\usepackage{array}
\usepackage{url}
\usepackage{float}
\usepackage{bm}
\usepackage{adjustbox}

\usepackage{subcaption}
\usepackage{diagbox}

\usepackage{amsthm}
\usepackage{tabularx}
\usepackage{mathrsfs} 
\usepackage{soul}

\usepackage{stfloats}
\usepackage[colorlinks=true, linkcolor=blue, citecolor=blue]{hyperref}

\usepackage{cleveref}
\begin{document}

\def\BibTeX{{\rm B\kern-.05em{\sc i\kern-.025em b}\kern-.08em
    T\kern-.1667em\lower.7ex\hbox{E}\kern-.125emX}}

    %Physical Layer Challenge-Response Authentication for Ambient Backscatter Devices
%Physical Layer Authentication between Ambient Backscatter Devices Based on Challenge and Response

\title{Physical Layer Challenge-Response Authentication between Ambient Backscatter Devices}

\author{Yifan~Zhang,~\IEEEmembership{Graduate Student Member,~IEEE,}
        Yongchao Dang,~\IEEEmembership{}
        Masoud Kaveh\IEEEmembership{},\\
        Zheng~Yan,~\IEEEmembership{Fellow,~IEEE,}
        Riku Jäntti,~\IEEEmembership{Senior Member,~IEEE,}
        and Zhu~Han,~\IEEEmembership{Fellow,~IEEE,}
\thanks{Y. Zhang, Y. Dang, M. Kaveh, and J. Riku are with the Department of Information and Communications Engineering, Aalto University, Espoo, 02150, Finland. (email: yifan.1.zhang@aalto.fi, masoud.kaveh@aalto.fi, yongchao.dang@aalto.fi, riku.jantti@aalto.fi)}
\thanks{Z. Yan (corresponding author) is with the State Key Lab of ISN, School of Cyber Engineering, Xidian University, Xi'an, Shaanxi, 710026 China. (email: zyan@xidian.edu.cn)}% 
\thanks{Z. Han is with the Department of Electrical and Computer Engineering, University of Houston, Houston, TX 77004 USA, and also with the Department of Computer Science and Engineering, Kyung Hee University, Seoul 446-701, South Korea (email: hanzhu22@gmail.com)}
}

% The paper headers
\markboth{Journal of \LaTeX\ Class Files}%
{Shell \MakeLowercase{\textit{et al.}}: A Sample Article Using IEEEtran.cls for IEEE Journals}

\maketitle

% and randomness of wireless channel fading. In the authentication procedure, challenge-response signals are exchanged at the physical layer, which allows two BDs to verify their pre-shared identity information while not revealing the information to attackers. 

\begin{abstract}
Ambient backscatter communication (AmBC) has become an integral part of ubiquitous Internet of Things (IoT) applications due to its energy-harvesting capabilities and ultra-low-power consumption. However, the open wireless environment exposes AmBC systems to various attacks, and existing authentication methods cannot be implemented between resource-constrained backscatter devices (BDs) due to their high computational demands.
To this end, this paper proposes PLCRA-BD, a novel physical layer challenge-response authentication scheme between BDs in AmBC that overcomes BDs’ limitations, supports high mobility, and performs robustly against impersonation and wireless attacks.
It constructs embedded keys as physical layer fingerprints for lightweight identification and designs a joint transceiver that integrates BDs' backscatter waveform with receiver functionality to mitigate interference from ambient RF signals by exploiting repeated patterns in OFDM symbols.
Based on this, a challenge-response authentication procedure is introduced to enable low-complexity fingerprint exchange between two paired BDs leveraging channel coherence, while securing the exchange process using a random number and unpredictable channel fading. Additionally, we optimize the authentication procedure for high-mobility scenarios, completing exchanges within the channel coherence time to minimize the impact of dynamic channel fluctuations. Security analysis confirms its resistance against impersonation, eavesdropping, replay, and counterfeiting attacks. Extensive simulations validate its effectiveness in resource-constrained BDs, demonstrating high authentication accuracy across diverse channel conditions, robustness against multiple wireless attacks, and superior efficiency compared to traditional authentication schemes.

\end{abstract}


\section{Introduction}
\newcommand{\minitab}[2][l]{\begin{tabular}{#1}#2\end{tabular}}

Ambient backscatter communication (AmBC) is a key enabler for the Internet of Things (IoT) and a promising technology for 6G massive communication, thanks to its energy-efficient and sustainable design \cite{8999426, 10195838}. By harvesting energy from the environment and reflecting incident ambient RF signals (e.g., Wi-Fi, TV signals), backscatter devices (BDs) enable uplink communication to dedicated readers and device-to-device (D2D) interactions without additional energy consumption for radio frequency (RF) transmission \cite{liu2013ambient,parks2014turbocharging}. However, the broadcast nature of AmBC exposes the backscatter signals from BDs to be easily intercepted or manipulated, leaving systems vulnerable to impersonation \cite{6907930} and wireless spoofing attacks \cite{shan2013phy}. In particular, this threat becomes even more pressing in distributed AmBC scenarios where resource-constrained BDs interact directly without a central reader’s oversight. Therefore, implementing effective and robust BD-to-BD authentication is critical to safeguarding the integrity of AmBC and ensuring secure, reliable connectivity.

There are two potential approaches to enhancing authentication for AmBC systems: cryptography methods and physical layer authentication (PLA). Traditional cryptographic approaches, which rely on pre-shared identity keys and complex encryption algorithms \cite{chien2007mutual, wang2012server, cho2015consideration, gao2019secure}, demand significant computational resources, making them unsuitable for resource-limited BDs. In addition, while lightweight cryptographic protocols \cite{chien2007sasi, 1599066, sun2009security} reduce complexity through bitwise operations or cyclic shifts, they fail to provide robust security under various attacks. In contrast, PLA leverages unique device-specific or channel-specific RF characteristics as device fingerprints for lightweight authentication, eliminating the consumption of key management and data encryption. Device-based fingerprints \cite{RN176,geneprint, RCID,2019RF-Mehndi,HUFU} exploit intrinsic hardware features, while channel-based fingerprints \cite{zanetti2010physical,BatAu,chang2024apauth,BCAuth,yang2024batchauth} rely on uncorrelated spatial features to distinguish legitimate devices from attackers. These features position PLA as a feasible solution for lightweight and secure authentication in AmBC systems.

Despite their potential, existing PLA schemes still face significant challenges, particularly in adapting to BD-to-BD AmBC scenarios, accommodating device mobility, and defending against eavesdropping attacks.  On the one hand, device-based fingerprints offer inherent uniqueness and high authentication accuracy \cite{RN176,geneprint,RCID,2019RF-Mehndi,HUFU}, yet extracting them demands sophisticated signal-processing techniques, which is impractical for BDs. Although these fingerprints are stable against environmental changes, they struggle with fast channel fading and Doppler shifts in mobile scenarios, and their unencrypted nature makes them susceptible to eavesdropping.
On the other hand, some schemes \cite{geneprint,BatAu} rely on stable channel characteristics, such as time of arrival and channel state information (CSI), to create unique fingerprints. However, these approaches depend on accurate channel estimation, which is an arduous task for BDs equipped with only simple demodulators \cite{parks2014turbocharging}. Moreover, these schemes require minimal channel variation within the coherence time, making them unsuitable for dynamic environments involving unmanned aerial vehicles (UAVs) or the Internet of vehicles (IoV).
Although recent studies design protocols that explore spatial fingerprints \cite{BCAuth} and joint CSI \cite{yang2024batchauth} to support mobile BDs, they still face challenges under high device mobility, remain vulnerable to eavesdropping, and depend on channel estimation. Thus, there is an urgent need for an authentication solution that can support BD-to-BD scenarios in AmBC systems, maintain effectiveness with device mobility, and ensure robust security against eavesdropping attacks.

In recent years, researchers have implemented the traditional challenge-response authentication framework \cite{chien2007mutual,
wang2012server,cho2015consideration,gao2019secure} at the physical layer, known as PL-CRA schemes \cite{shan2013phy,7501833,shoukry2015pycra}. These schemes leverage properties of the wireless medium (e.g., channel fading and noise) to secure key exchanges and utilize device-specific fingerprints embedded in response signals for authentication. In such schemes, a verifier first transmits a challenge signal to the target device, which may actively modulate its identity on the received challenge \cite{shan2013phy,7501833} or passively reflect it \cite{shoukry2015pycra} to respond to the verifier. The verifier then extracts identity features from the response using prior knowledge of the challenge and compares them with stored reference fingerprints for authentication. These methods have demonstrated effectiveness across diverse environments and offer resilience against eavesdropping \cite{shan2013phy}. However, they require the presence of powerful transceivers capable of active signal transmission and processing, rendering them unsuitable for authentication between two passive BDs. They also cannot be directly applied to ambient scenarios where an ambient RF source continuously transmits interference signals.

% although the physical-layer challenge-response authentication mechanism (PL-CRAM) first proposed in \cite{shan2013phy} supports high-mobility authentication and resists wireless attacks, it is designed for scenarios consisting of active transceivers and is thus incompatible with passive BDs. Furthermore, while \cite{7501833} incorporates artificial noise into PL-CRAM for OFDM systems to counter key-recovery attacks, and \cite{shoukry2015pycra} leverages environmental physics to protect active sensors from spoofing attacks, these approaches rely on complex active signal processing and cannot be directly applied to ambient scenarios where an ambient RF source continuously transmits interference signals. 

%For example, Shan et al. \cite{shan2013phy} proposed PL-CRAM, a physical layer authentication mechanism where the authenticated device performs inverse operations for challenge signals and modulates its own key to generate a response signal that can eliminate channel fading between the two devices. Meanwhile, they eliminate pilot signals to prevent attackers from decoding the exchanged messages by channel estimation.

To address these challenges, we propose PLCRA-BD, a physical layer challenge-response authentication scheme designed explicitly between BDs in AmBC systems. PLCRA-BD uses embedded keys as unique fingerprints and designs a joint transceiver for mutual extraction while mitigating ambient interference. It also includes a challenge-response physical-layer authentication procedure that uses ambient RF signals for key exchange between passive BDs, eliminating active RF transmission and complex processing. Low-complexity operations and channel coherence ensure secure, mobility-resilient authentication, with security analysis confirming its robustness in dynamic AmBC environments. Specifically, the main contributions of this paper are summarized as follows:
%To this end, we proposed the physical-layer challenge-response authentication mechanism between BDs for the AmBC system. First, embedded keys are designed and stored in the memory of each BD as unique fingerprints. These keys are encoded as harvested power values, making them easily extractable by the BD. Second, a joint transceiver is developed to integrate the BD’s backscatter waveform with its receiver functionality, enabling BDs to extract each other’s fingerprints while effectively mitigating interference from ambient RF sources.
%Third, unlike traditional  PL-CRAM, the proposed authentication mechanism utilizes ambient OFDM signals to facilitate key exchange between passive BDs, eliminating the need for active RF transmission and complex signal processing. Additionally, it employs low-complexity operations and leverages channel coherence to ensure secure key exchange without relying on channel estimation or signal decoding, thereby minimizing the impact of mobility on authentication.Security analysis and performance evaluation demonstrate the robustness and reliability of the proposed mechanism, safeguarding the authentication process in dynamic AmBC environments. The main contributions of this paper are summarized as follows:

\begin{itemize}
\item \textbf{Embedded Fingerprint Construction:} To facilitate low-complexity fingerprint transmission and verification, we design an embedded fingerprint construction method for BDs. This approach replaces traditional device features with low-density physical-layer identity (PID) keys stored in the memory unit of each BD as its unique fingerprint. These PID keys are transmitted using simple amplitude shift keying (ASK) modulation, allowing receiver BDs to easily extract the PID-bearing signals through harvested power detection. Additionally, the PID keys are shared among BDs to facilitate lightweight fingerprint verification.

  %for efficient message exchange between BDs

\item \textbf{Joint Transceiver Design:}  
To mitigate ambient interference in AmBC systems, a joint transceiver design integrates the backscatter waveform of BDs with their receiver functionality. By inserting an amplitude hop at the midpoint of the backscatter waveform and exploiting the repeated cyclic prefix structure of downlink orthogonal frequency division multiplexing (OFDM) signals, the receiving BD can isolate the backscatter signal from superposed components. This approach effectively avoids interference from the downlink signal and enables efficient message exchange between BDs.

%This is achieved by analyzing the superposed signal in AmBC, which comprises signals backscattered by other BDs and signals directly from the ambient source. This design enables BDs to extract the fingerprints while effectively mitigating interference from ambient RF sources. 
%Challenge-response authentication procedure between BDs in AmBC
\item \textbf{One-way and Mutual Authentication:} Building on embedded fingerprint and integrated transceiver design, we propose a challenge-response authentication procedure to achieve one-way and mutual authentication between BDs in AmBC systems, ensuring high mobility and robust security. In one-way authentication, a verifier BD challenges a prover BD with a random number and its PID key within the channel coherence time. The prover mitigates channel effects by dividing the harvested power of the two received signals and uses the verifier's PID key to estimate the random number. The prover then responds with the estimated random number and its PID key. The verifier, in turn, estimates the prover's PID using a similar method and authenticates it accordingly. For mutual authentication, the process is repeated bidirectionally between the BDs. This method eliminates channel estimation and minimizes CSI fluctuations by exploiting channel coherence, as well as preserves shared PID key secrecy leveraging dynamically generated random numbers and unpredictable channel fading.

\item \textbf{Security Analysis and Comprehensive Performance Evaluation:} We theoretically analyze the resistance of PLCRA-BD to impersonation, eavesdropping, replay, and counterfeiting attacks. Comprehensive simulation results further demonstrate the desirable performance of PLCRA-BD across key performance metrics, including effectiveness, authentication accuracy, robustness, and efficiency. The effectiveness of the scheme is validated through a proof-of-concept experiment and theoretical complexity analysis. High authentication accuracy is achieved under diverse channel conditions, such as varying speeds and distances between BDs. Robustness is evaluated under various wireless attacks. Furthermore, the scheme exhibits exceptional efficiency, characterized by low latency and minimal power consumption. Comparative evaluations with traditional schemes highlight the scheme’s superior performance in most evaluated dimensions.

\end{itemize}

The remainder of this paper is organized as follows: Section \ref{related work} reviews related work. Section \ref{System Setup} describes the system model and outlines the problem. Section \ref{AuthScatter} details the PLCRA-BD design, and Section \ref{Security Analysis} analyzes the security of PLCRA-BD. Section \ref{Evaluation} evaluates the performance of PLCRA-BD, and Section \ref{Conclusion} concludes the paper.

\begin{table*}[]
\footnotesize
\centering
\caption{Comparison PLCRA-BD with existing PLA works in BC systems.}
\label{related_work}
\vspace{1mm}
{\fullcirc: satisfy a criterion; \emptycirc: do not satisfy a criterion; \halfcirc: partly satisfy a criterion.}
\\[2mm]

\begin{tabular}{c|c|c|c|cc|cccc}
\toprule[1.5pt]
\multirow{2.5}{*}{Classification} & \multirow{2.5}{*}{Works} & \multirow{2.5}{*}{\minitab[c]{AmBC \\ support}} & \multirow{2.5}{*}{\minitab[c]{Authentication\\between BDs}} & \multicolumn{2}{c|}{Mobility} & \multicolumn{4}{c}{Attack Resistance} \\ \cmidrule{5-6} \cmidrule{7-10}
                       &             &        & & Device & Environment & IA & EA & RP & CA \\ \midrule

\multirow{4}{*}{\minitab[c]{Device\\ Fingerprint}} 

& Harvestprint \cite{RN176} & \emptycirc & \emptycirc & \emptycirc & \fullcirc & \fullcirc & \emptycirc & \emptycirc & \emptycirc \\
&Geneprint \cite{geneprint} & \emptycirc & \emptycirc & \emptycirc & \fullcirc & \fullcirc & \emptycirc & \fullcirc & \emptycirc \\
& RCID \cite{RCID}  & \emptycirc & \emptycirc & \emptycirc & \fullcirc & \fullcirc & \emptycirc & \fullcirc & \emptycirc \\
& RF-Mehndi \cite{2019RF-Mehndi}  & \emptycirc & \emptycirc & \emptycirc & \fullcirc & \fullcirc & \emptycirc & \fullcirc & \emptycirc \\ 

& Hu-Fu \cite{HUFU} & \emptycirc & \emptycirc & \emptycirc & \fullcirc & \fullcirc & \emptycirc & \fullcirc & \emptycirc \\\midrule

\multirow{5}{*}{\minitab[c]{Channel\\ Fingerprint}} 
 & Zanetti et al. \cite{zanetti2010physical} & \emptycirc & \emptycirc & \emptycirc & \emptycirc & \fullcirc & \emptycirc & \emptycirc & \emptycirc \\

% & Butterfly \cite{butterfly} & \emptycirc & \emptycirc & \emptycirc & \fullcirc & \fullcirc & \halfcirc & \fullcirc & \fullcirc \\
& BatAu \cite{BatAu} & \emptycirc & \emptycirc & \emptycirc & \fullcirc & \fullcirc & \emptycirc & \fullcirc & \emptycirc \\
& APAuth \cite{chang2024apauth} & \emptycirc & \emptycirc & \emptycirc & \fullcirc & \fullcirc & \emptycirc & \fullcirc & \fullcirc \\ 
& BCAuth \cite{BCAuth} & \emptycirc & \emptycirc & \halfcirc & \fullcirc     & \fullcirc & \emptycirc & \fullcirc & \fullcirc \\

& BatchAuth \cite{yang2024batchauth} & \emptycirc & \emptycirc & \halfcirc & \fullcirc & \fullcirc & \emptycirc & \fullcirc & \fullcirc \\

\midrule
\multicolumn{2}{c|}{PLCRA-BD} & \fullcirc & \fullcirc & \fullcirc & \fullcirc & \fullcirc & \fullcirc & \fullcirc & \fullcirc \\

\bottomrule[1.5pt]
\end{tabular}

\begin{tablenotes}
        \item \hspace{10mm} 1. IA: impersonation attacks; EA: eavesdropping attacks; RP: replay attacks; CA: counterfeiting attacks.
        \item \hspace{10mm} 2. The results presented in the table are derived from the original results in the corresponding paper.
    \end{tablenotes}
% \vspace{-2mm}
\end{table*}

\section{Related Works} \label{related work}
This section reviews PLA schemes in BC systems and related PL-CRA schemes. Table \ref{related_work} highlights that the existing PLA schemes in BC systems lack BD-to-BD authentication in AmBC scenarios, full mobility support, and robust resistance to eavesdropping attacks, which our PLCRA-BD addresses.

\subsection{Physical Layer Authentication in BC}
Two types of RF fingerprints are mainly exploited by existing PLA schemes in BC systems: device fingerprints and channel fingerprints.

%Table~\ref{related_work} compares our work with these PLA schemes and shows that the existing literature still lacks a scheme that can enable authentication between BDs in an AmBC system, fully support device mobility, and perform robustly against various wireless attacks. 

\subsubsection{Device Fingerprints} 
Device fingerprints are derived from inherent hardware variations caused by random deviations during the manufacturing process. 
For example, Narayanan et al. \cite{RN176} exploited unique energy discharge patterns in BD oscillators as fingerprints to prevent impersonation attacks.  Han et al. \cite{geneprint} proposed Geneprint, leveraging covariance and power spectrum density similarities between successive backscattered signals as fingerprints, combined with supervised learning for BD identification.  In a system called RCID \cite{RCID}, the reflection coefficient of a BD is captured as the fingerprint, which is then differentiated by training a multi-class neural network. RF-Mehnd \cite{2019RF-Mehndi} exploited the phase shift caused by the human touching of a BD array as an authentication fingerprint for both the BD and its holder. Then, a classifier is developed that uses a support vector machine to validate the fingerprint. Han et al. proposed Hu-Fu \cite{HUFU}, which uses RF signal differences between BD pairs, leveraging power spectral density, energy spectrum, and a cross-correlation-based threshold for authentication.

The aforementioned schemes are capable of supporting environmental mobility, as hardware features are insensitive to changes in the surrounding environment. Most of these schemes \cite{geneprint,RCID,2019RF-Mehndi,HUFU} demonstrate resilience against impersonation and replay attacks due to the uniqueness of device fingerprints. However, they generally overlook device mobility and are unable to prevent eavesdropping or counterfeiting attacks from learning the constant features of the backscatter signal \cite{BCAuth}. Moreover, these schemes often require a complex analyzer on the verifier side for fingerprint extraction and analysis, making them unsuitable for resource-constrained BDs. Additionally, all these schemes operate in monostatic configurations and fail to address BD to BD authentication in AmBC scenarios.

\subsubsection{Channel Fingerprints}
Channel fingerprinting, using unique wireless signal characteristics like fading, reflection, and scattering, is widely studied in existing BC systems for authentication. Zanetti et al. \cite{zanetti2010physical} utilized average baseband power and time interval errors of backscatter signals to counter impersonation attacks, but their approach suffers from low accuracy of fingerprint extraction in dynamic environments. Yang et al. \cite{BatAu} proposed a batch authentication scheme using joint CSI with PD-NOMA techniques, effectively resisting impersonation and replay attacks while adapting to environmental mobility; however, its complex fingerprint extraction process limits its applicability for BDs. To simplify this, Chang et al. \cite{chang2024apauth} introduced APAuth, enabling BDs to authenticate readers via harvested power volumes without complex operations. While lightweight, APAuth relies on pre-negotiated power levels, supports only reader-to-BD authentication, and does not address BD-to-BD scenarios. Additionally, the above schemes face challenges such as vulnerability to eavesdropping and counterfeiting attacks due to unencrypted signals and the open nature of wireless channels, their restriction to monostatic scenarios unsuitable for AmBC systems, and their limited consideration of device mobility, reducing their effectiveness in dynamic environments.

%. Then, the similarity between the extracted fingerprint and the real fingerprint is measured by Euclidean distance, which is compared with an experience-trained threshold for authentication, which can prevent impersonation attacks. 
% Based on that,  Yang et al. \cite{yang2024batchauth} further exploited the channel correlation coefficient to make an authentication decision for mobile devices and theoretically prove the robustness of their scheme when facing counterfeiting attacks. the fingerprint they used can be easily eavesdropped and replicated in an open wireless channel. be used when BDs are moving, which causes fast changes in the required CSI at the verifier. These two methods cannot

Recent works \cite{BCAuth,yang2024batchauth} have designed robust authentication scheme for mobile BC scenarios and against wireless attacks on BDs. In BCAuth \cite{BCAuth}, the authors utilized the spatial information of a BD as its fingerprint and employed a tracing algorithm to update the fingerprint dynamically when the BD moves. BatchAuth \cite{yang2024batchauth} leveraged joint CSI to authenticate multiple BDs simultaneously, adapting to CSI variations of mobile BDs through channel correlation coefficients, which further improves the authentication efficiency for mobile BDs. Both BCAuth and BatchAuth support authentication for moving BDs but struggle in high-speed scenarios with rapid CSI changes. Moreover, fingerprints based on device location \cite{BCAuth} and CSI \cite{yang2024batchauth} are still susceptible to eavesdropping and targeted counterfeiting based on the eavesdropping results in open wireless environments. Furthermore, these authentication protocols rely on channel estimation, which is incompatible with BDs due to their limited computational and communication capabilities. More seriously, BDs in AmBC face challenges in isolating ambient signals for effective fingerprint extraction. Thus, an urgent need exists for a solution that enables authentication between BDs in AmBC systems, addresses mobility challenges, and ensures robust security.

% In such a way, they can authenticate moving BDs and prevent attackers from imitating the fingerprints due to the uniqueness of the spatial feature. 
\begin{figure}[tp]
    \centering
    \includegraphics[width=\linewidth]{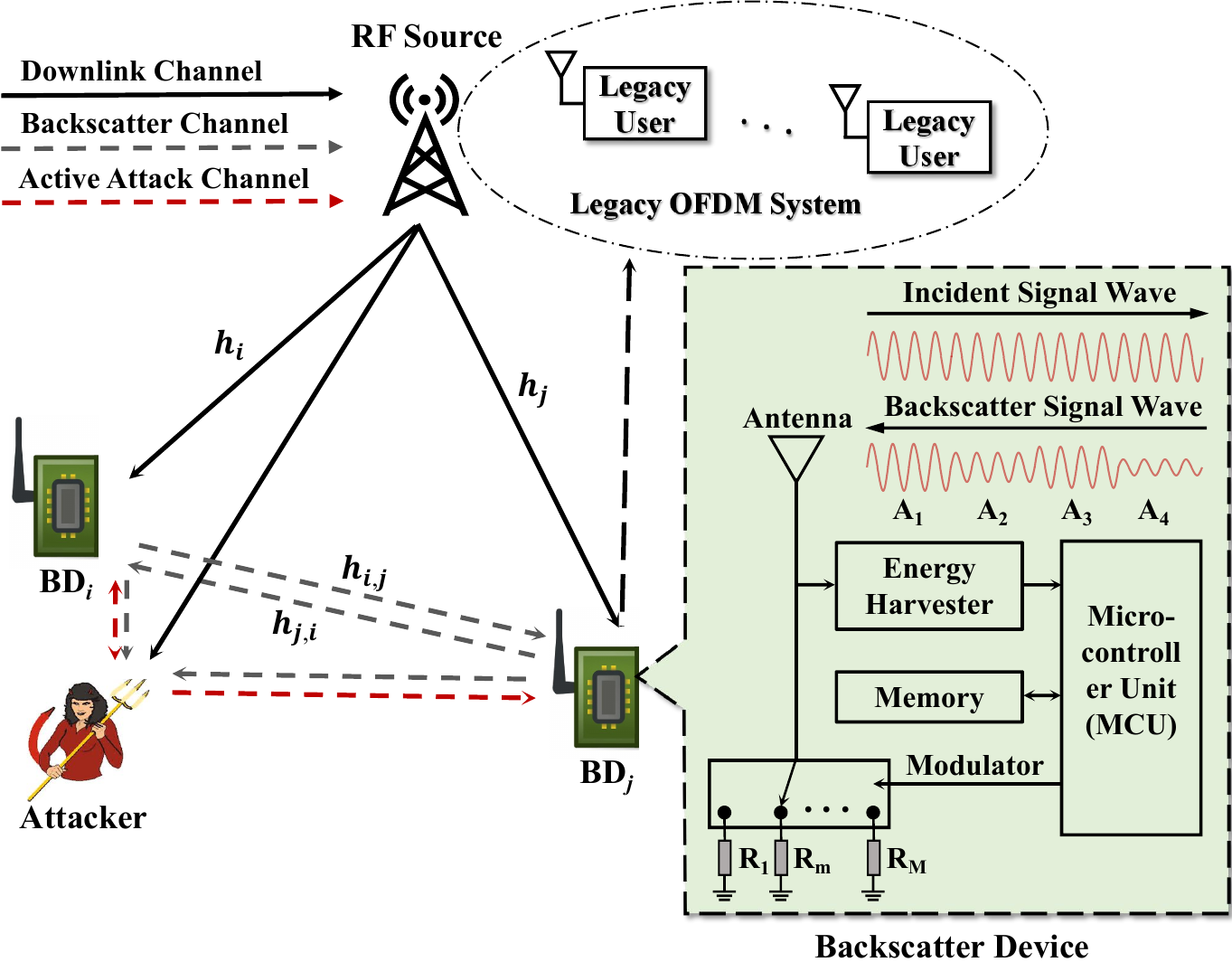}
    \caption{System model for AmBC using ambient
RF signals.}
    \label{syse}
\end{figure}

%where the authenticated device performs inverse operations for challenge signals and modulates its own key to generate a response signal that can eliminate channel fading between the two devices. Meanwhile, they eliminate pilot signals to prevent attackers from decoding the exchanged messages by channel estimation.

\subsection{PL-CRA}
Several PL-CRA schemes have been proposed to achieve challenge-response authentication at the physical layer \cite{shan2013phy,du2014physical,7501833,shoukry2015pycra}. For example, Shan et al. \cite{shan2013phy} first proposed PL-CRAM, a physical layer authentication mechanism, which eliminates pilot references to prevent attackers from estimating legitimate channels, while the verifier can decode the challenge-response messages without CSI by exploiting the former challenge signal to compensate for the channel fading in the latter response signal. 
%This solution was further extended to multi-hop networks in \cite{du2014physical}. 
Then, Wu et al. added artificial noise to protect the challenge-response process \cite{7501833}, and Shoukry et al. \cite{shoukry2015pycra} deployed active actuators to continuously challenge the surrounding environment with random transmissions for attacker detection.
The above-mentioned schemes can satisfy mobility and defend against various attacks. However, they require the verifier to emit challenge signals actively and perform signal processing to extract the challenge-response messages, which cannot apply to passive BDs due to their limited power and processing abilities. In addition, they neglect the ambient scenarios where an uncontrollable ambient RFS continuously broadcasts ransom signals, which could face difficulties when deploying in AmBC systems.

To this end, we introduce the PLCRA-BD. Unlike existing PL-CRA schemes, PLCRA-BD utilizes ambient OFDM signals to facilitate key exchange between passive BDs, eliminating the need for active RF transmission and complex signal processing. Additionally, it employs low-complexity operations and leverages channel coherence to ensure secure key exchange without relying on channel estimation or signal decoding, thereby minimizing the impact of mobility on authentication. In the next section, we describe the system model and state the problem that needs to be solved in this paper.

% First, embedded keys are designed and stored in the memory of each BD as unique fingerprints. These keys are encoded as harvested power values, making them easily extractable by the BD. Second, a joint transceiver is developed to integrate the BD’s backscatter waveform with its receiver functionality, enabling BDs to extract each other’s fingerprints while effectively mitigating interference from ambient RF sources.
% Third, unlike traditional CR-PLAM, the proposed authentication mechanism utilizes ambient OFDM signals to facilitate key exchange between passive BDs, eliminating the need for active RF transmission and complex signal processing. Additionally, it employs low-complexity operations and leverages channel coherence to ensure secure key exchange without relying on channel estimation or signal decoding, thereby minimizing the impact of mobility on authentication. Security analysis and performance evaluation demonstrate the robustness and reliability of the proposed mechanism, safeguarding the authentication process in dynamic AmBC environments. In the next section, we statement the problem that

\section{System Model and Problem Statement} \label{System Setup}

\subsection{System Model} \label{system model}
As depicted in Fig.~\ref{syse}, the AmBC system consists of an ambient RF source (e.g., TV tower, cellular base station, or Wi-Fi transmitter) and two paired BDs, denoted as $BD_i$ and $BD_j$. The RF source operates using a legacy OFDM system to transmit signals to its users, while the BDs communicate either with each other or with legacy users by reflecting the RF source’s signals. Communication in the AmBC system follows a time-division structure, ensuring that each BD reflects the ambient RF signals exclusively within its assigned time slot, thereby avoiding inter-BD interference and maintaining efficient operation. In addition, the notations used in this paper are presented in Table \ref{notations}.

Each BD, equipped with a single antenna, consists of an energy harvester, a backscatter modulator, a memory module, and a microcontroller unit (MCU)—a standard configuration for passive backscatter devices \cite{parks2014turbocharging,liu2013ambient}. BDs operate in two primary modes: (i) backscattering, where the BD reflects incident signals and uses $M$ adjustable impedances to implement M-ary amplitude shift keying (M-ASK) modulation \cite{goay2024optimal}, and (ii) listening, where the BD switches its antenna to harvest energy from incoming signals. The memory unit enables the storage of simple number sequences \cite{galisteo2020two}, such as the identity sequence of devices, while the MCU facilitates simple computations, such as multiplications, to enable authentication and streamline system operations.

\begin{table}[!t]
\caption{List of notations used in \textsc{PLCRA-BD}}
\centering
\begin{tabular}{@{}ll@{}}
\toprule
\textbf{Notation} & \textbf{Description} \\ \midrule
$\mathcal{CN}(\mu, \sigma^2)$ & circularly symmetric complex Gaussian (CSCG) distribution. \\ %with mean $\mu$ and variance $\sigma^2$
$|\cdot|$ & absolute value of a scalar or a complex number. \\
$||\cdot||_{l1}$ & $L_1$-norm, sum of the absolute values of a vector. \\
%$>$, $<$ & Greater-than and less-than comparison operators, respectively. \\
$\min(\cdot)$, $\max(\cdot)$ & minimum and maximum functions for a given set. \\
$O(L)$ & asymptotic time complexity on the order of $L$. \\
$P(s(t))$ & average power of the signal $s(t)$. \\
%, typically defined as $P(s(t)) = \frac{1}{T}\int_{0}^{T}|s(t)|^2 dt$
$\eta_i$ & energy harvesting efficiency at the $i$-th device. \\
\bottomrule
\end{tabular}
\label{notations}
\end{table}

\subsection{Channel Model in AmBC Systems} \label{Signal Model}

The downlink channels from the RF source to $BD_i$, $BD_j$, and an attacker are represented as $h_i(t)$, $h_j(t)$, and $h_a(t)$, respectively. The inward channels between $BD_i$, $BD_j$, and the attacker are denoted as $h_{l,k}(t)$ with $l,k \in \{i,j,a\}$. A channel gain can be expressed as $h = \vartheta d^{\lambda/2}$, where $\vartheta$ is a CSCG variable, $d$ is the transmitter-receiver distance, and $\lambda$ is the path-loss exponent. Let \( s(t) \) denote the signal transmitted from the RF source, which is usually OFDM-based for existing ambient RF sources, such as TV tower and Wi-Fi router. \( b_i(t) \) denotes the backscatter signal of \( BD_i \), which represents the reflection coefficient at $BD_i$ in the M-ASK case. 
% a bandpass signal transmitted from the RF source during a symbol interval as
% \begin{equation}
%     \tilde{s}(t) = \Re\{\sqrt{p}s(t)e^{j2\pi f_c t}\},
% \end{equation}
% where \( s(t) \) is a unit baseband signal with transmission power \( p \), and \( f_c \) represents carrier frequency. The ambient signal received at \( BD_i \) can be represented as \cite{liu2013ambient}
% \begin{equation}
%     \tilde{c}_i(t) = \Re\left\{\left[\sqrt{p }h_{i}(t) s(t)\right] e^{j2\pi f_c t}\right\}, \tag{2}
% \end{equation}
% where \( c_i(t) = \sqrt{p }h_{R,B_i}(t) s(t) \) is the baseband representation of \(\tilde{c}_i(t)\).
%The signal backscattered from \( B_i \) is \( \tilde{c}_i(t) b_i(t) \). 
Thus, the received superposed signal at \( BD_j \) that contains the backscatter signals from \( BD_i \) and the downlink signals directly from the ambient source can be expressed as

% \begin{equation}
%     {y}_j(t) = h_{i,j}(t)h_{i}(t) b_i(t) + h_{j}(t){s}(t) + {w}_j(t), \label{3}
% \end{equation}

\begin{equation}
\begin{split}
    {y}_j(t) &= h_{i,j}(t)h_{i}(t) b_i(t){s}(t) + h_{j}(t){s}(t) + {w}_j(t) \\
    &= y_j^b(t) + y_j^d(t) + w_j(t), \label{channel}
\end{split}
\end{equation}
% where \( {w}_j(t) \) is the received additive white Gaussian noise (AWGN) at \( A_j \). The baseband representation of (\ref{3}) can be siplified as:
% \begin{equation}
%     y_j(t) = y_j^b(t) + y_j^d(t) + w_j(t),  
% \end{equation}
where \( y_j^b(t) =   h_{i,j}(t)h_{i}(t) b_i(t){s}(t) \) is the backscatter signal reflected from \( BD_i \), \( y_j^d(t) =  h_{j}(t){s}(t) \) is the downlink ambient signal directly from the RF source, and \( {w}_j(t) \) is the received additive white Gaussian noise (AWGN) at \( BD_j \), \( w_j(t) \sim \mathcal{CN}(0, \sigma_j^2) \).

Let \( h_i[n] \), \( h_j[n] \) and \( h_{i,j}[n] \) denote the discrete-time representation of \( h_i(t) \), \( h_j(t) \) and \( h_{i,j}(t) \), respectively. The ambient signal is \( s[n] \), and the backscatter signal at $BD_i$ is $b_i[n]$. Thus, the representation in   (\ref{channel}) of the received signal at \( BD_j \) can be rewritten to a discrete-time form  as
\begin{equation} \label{5}
    y_j[n] = y_j^b[n] + y_j^d[n] + w_j[n],
\end{equation}
where \( y_j^b[n] =  h_{i,j}[n]h_{i}[n] b_i[n]s[n] \), \( y_j^d[n] =  h_j[n] s[n] \) and \( w_j[n] \sim \mathcal{CN}(0, \sigma_j^2) \).

\begin{figure}[tp]
    % \vspace{-0.2cm}
    \centering
    \includegraphics[width=\linewidth]{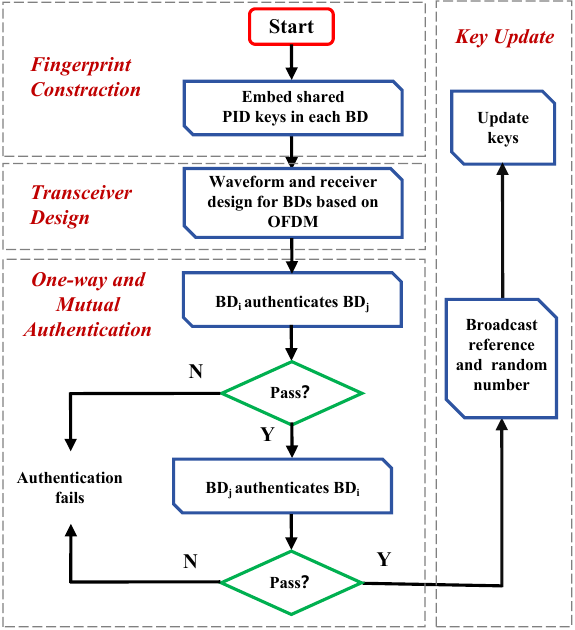}
    % \vspace{-0.1cm}
    \caption{PL-CRABD Overview}
    \label{Overview}
%\vspace{-0.6cm}
\end{figure}

\subsection{Adversary Model} \label{Adversary Model}

In this work, the attacker in Fig.~\ref{syse} is assumed to employ four types of attacks: one impersonation attack and three wireless attacks. The impersonation attack, operating at the link or network layer, involves using known upper-layer identifiers to masquerade as a legitimate BD. In contrast, the wireless attacks are assumed to be aware of the PLA procedure and attempt to manipulate the signals in the physical layer. Specifically, they are defined as follows:

\begin{itemize}
    \item \textit{Impersonation Attacks:} The attacker leverages known upper-layer identifiers (e.g., reference numbers) of a genuine device to pose as a legitimate BD, thereby attempting to pass off its transmissions as originating from an authorized entity.
    \item \textit{Eavesdropping Attacks:} Attackers intercept the signals exchanged between BDs to deduce their secret keys. We consider two types of eavesdroppers: a \textit{native eavesdropper} positioned beyond the coherence distance with a legitimate BD, and a \textit{smart eavesdropper} located within the coherence distance with a legitimate BD, thereby gaining more favorable conditions for signal interception.
    \item \textit{Replay Attacks:} Attackers record a legitimate transmission and retransmit it at a later time, attempting to impersonate a legitimate BD.
    \item \textit{Counterfeiting Attacks:} Attackers eavesdrop on and imitate the information exchanged between the two BDs during authentication, aiming to produce RF waveforms indistinguishable from those of legitimate devices and thereby deceive the verifier.
\end{itemize}

%\subsection{Design Goals}
% Existing BDs can only decode signals as a series of 0-1 bits by detecting power fluctuations, leaving transmitted messages vulnerable to attacks. Additionally, achieving PLA in BDs is challenging due to their limited ability to extract RF features and eliminate interference from downlink ambient signals.

In this paper, we aim to propose a PL-CRABD scheme for authentication between BDs in the AmBC system, designed to support mobility and provide high robustness against impersonation and wireless attacks. To facilitate authentication, the scheme should address BDs' limitations in extracting fingerprints and mitigating interference from ambient downlink signals. For mobility support, the scheme should operate effectively with mobile BDs at various speeds and in diverse environments, including rural and urban scenarios. To ensure robustness, the scheme should perform robustly under various wireless attack scenarios. The next section presents the details of the proposed PL-CRABD design.

\section{PL-CRABD Design} \label{AuthScatter}
In this section, we present the detailed design of the proposed PL-CRABD scheme. We first present a brief overview and then introduce the key components of the scheme.

\subsection{Overview}
As shown in Fig. \ref{Overview}, the proposed PL-CRABD scheme comprises five key components: fingerprint construction, transceiver design, one-way and mutual authentication, and key update. In the fingerprint construction, each BD is assigned a unique PID key that serves as its fingerprint. These keys are shared among BDs that can be transmitted vai simple ASK modulation.
Next, a joint transceiver design enables BDs to extract the harvested power value of the backscattered signal while avoiding downlink interference. This is achieved by leveraging the CP repetition pattern in OFDM signals. Specifically, by introducing a controlled amplitude transition at the midpoint of the backscatter waveform, the receiver BD isolates the harvested backscattered signal power using the repeated CP pattern.
Based on the fingerprint construction and the transceiver design, a one-way authentication procedure is further proposed that consists of a challenge and a response stage.
In the challenge stage, the verifier BD transmits a random number and its PID key to the prover BD via backscattered ambient signals. Extracting the harvested power value of the two signals, the prover BD eliminates the channel fading effect using a simple division to derive a factor containing the random number and the verifier's PID. Using the stored PID key of the verifier, the prover calculates the random number value and backscatters the random number along with its own PID key to the verifier in the response stage. Using a similar computation, the verifier estimates the prover BD’s PID key and compares it with the stored key to confirm authentication. After that, Mutual authentication is achieved by repeating the one-way authentication procedure, allowing both BDs to authenticate each other in a secure and efficient manner.
Finally, to enhance long-term security, both BDs broadcast their reference and newly generated random numbers. Thus, all BDs can find the target BD and update their shared PID keys dynamically using the random number, ensuring continued protection against potential security threats.

\subsection{Fingerprint Construction}
We embed PID keys in the memory of BDs as fingerprints to facilitate low-complexity fingerprint exchange and extraction. These keys are stored as fixed values and can be modulated by ASK, allowing the signals carrying the key information to be obtained by reading the harvested power of the received signals. In addition, they are shared between BDs for identity verification. For example, the key for $BD_i$ is denoted as $|K_{  i}|$. For an AmBC system with $N$ BDs, total $N$ keys ${|K_1|, |K_2|, \ldots, |K_N|}$ are stored in each BD for authentication. $BD_i$ can display its key by backscattering a signal $K_{  i}$ with amplitude $|K_{  i}|$. Then, the receiver BD (e.g., $BD_j$) can obtain a key-bearing signal that contains the key from its harvested power value for the signal $K_{  i}$, denoted as $ \eta P(s(K_{  i}))$. By constructing the shard key as the fingerprint, $BD_{i}$ and $BD_{j}$ can perform authentication procedures to obtain each other's key-bearing signal without needing signal decoding.

%where $\eta$ represents the energy harvesting efficiency at the receiver BD and $P(\cdot)$ represents the calculation of the average power of a signal
\begin{figure}[tp]
    % \vspace{-0.2cm}
    \centering
    \includegraphics[width=\linewidth]{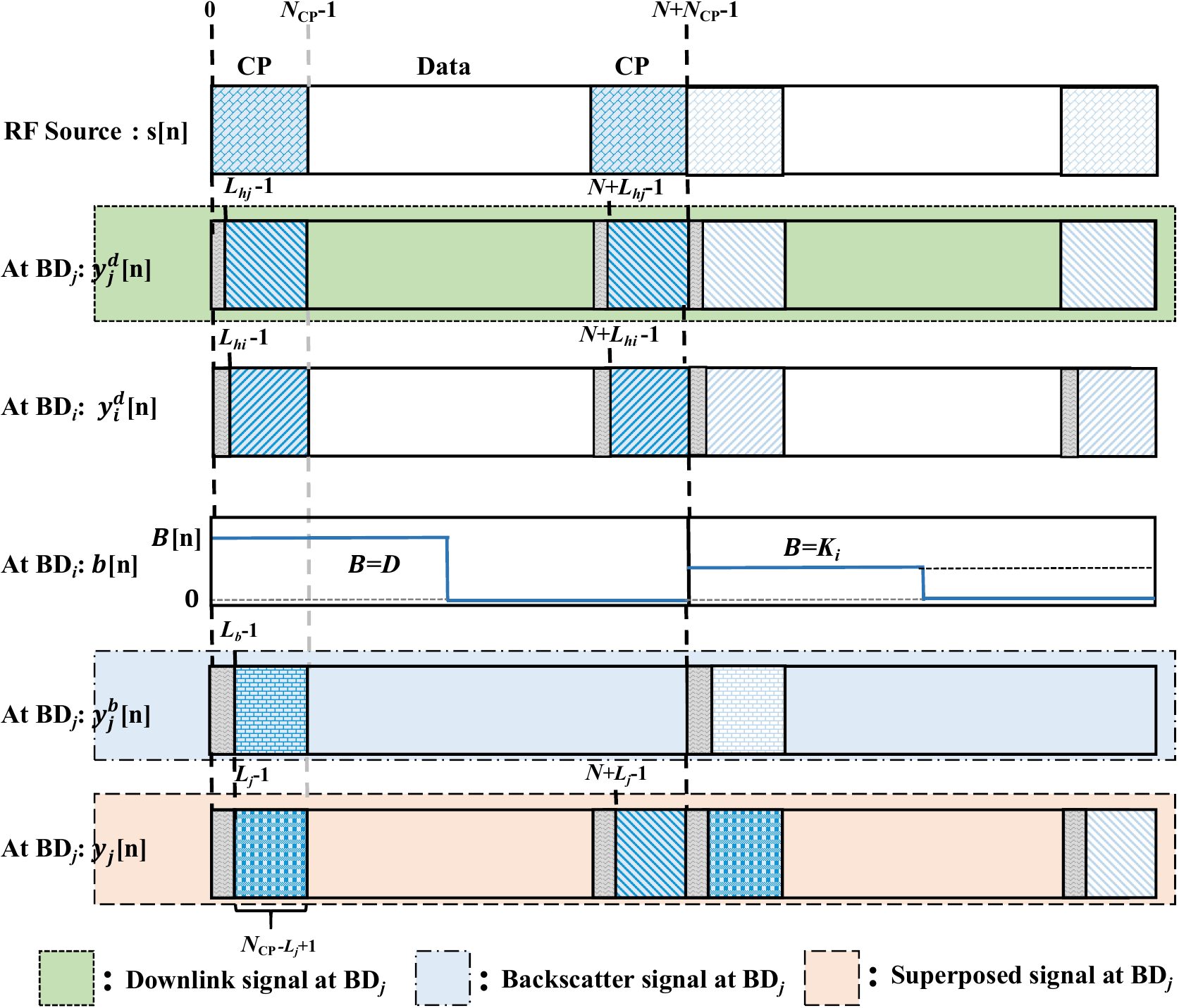}
    % \vspace{-0.1cm}
    \caption{OFDM signals at different stages when $L_j = L_{b_j}$, one OFDM symbol for $B = D$ and another for $B = K$. The superposed signals received at $BD_j$ is $y_j[n]=y^b_j[n]+y^d_j[n]$}
    \label{transceiver}
%\vspace{-0.6cm}
\end{figure}

\subsection{Transceiver
Design Over OFDM Carrier} \label{Enhancement}
%To let the receiver BD estimate the harvested power value only from the backscatter signal of another BD without the interference of the downlink signal from the RF source, we design a joint transceiver that integrates the BD backscatter waveform and BD receiver using the repeated pattern in the OFDM signal. For easy understanding, we present the design with a discrete-time AmBC system.

%The OFDM transmission scheme is widely used by the RF source in current ambient backscatter systems, such as LET \cite{8809924}, cognitive radio \cite{8672817}, and WiFi systems \cite{yang2023universal}.  In OFDM signals, a cyclic prefix (CP) is used to extend an OFDM symbol by copying the last samples of the OFDM symbol into its beginning so as to avoid interblock interference (ISI) caused by the multi-path delay spread of wireless channels \cite{proakis2008digital}. By exploiting this repeated CP pattern, we propose a joint transceiver design, including the BD backscatter waveform and the receiver, which enables a BD to extract the harvest power value from the backscatter signal of other BD without the effect of the signal from the ambient source. This design is similar to the direct-link interference cancellation mechanism \cite{yang2017modulation}, which mainly focuses on signal demodulation in a commonly used backscatter receiver, e.g., a reader. However, we focus on the D2D transmission between two BDs and the power domain of the signal. %Finally, we discuss the synchronization problem in our proposed transceiver design.

For clarity, we present the design in a discrete-time AmBC system. The OFDM signal 
\( s[n] \) includes CP that copies the tail of each OFDM sample to the start to mitigate inter-symbol interference (ISI) in multipath channels \cite{proakis2008digital}. Let the downlink channel be \( h_i[n] \)  with delay spreads \(\tau_{h_i}\) and the inward channel be \( h_{i,j}[n] \) with delay spreads \(\tau_{h_{i,j}}\), 
and let \( f_s \) be the OFDM sampling rate. Thus, 
\( L_{h_i} = \lceil \tau_{h_i} f_s \rceil \) and 
\( L_{h_{i,j}} = \lceil \tau_{h_{i,j}} f_s \rceil \). Define 
\( L_{b_j} = L_{h_i} + L_{h_{i,j}} \) and 
\( L_j = \max \{L_{h_j}, L_{b_j}\} \). Let \( N \) be the number of subcarriers, and let the CP length 
\( N_{cp} \) exceed the maximum channel spread, i.e., 
\(\tau_{j} \leq T_{\text{CP}}\). Without loss of generality, consider \( BD_i \) sending  a message to \( BD_j \).

% \vspace{-0.2cm}
\subsubsection{BD Backscatter Waveform Design}\label{Re-Authentication}

A backscatter waveform of BDs is designed in this subsection to make the backscatter signal arrive at a receiver that is distinguishable from the downlink signal. Specifically, in the backscattering mode, $BD_i$ backscatters the OFDM signal to transmit information, where the duration of each BD symbol is equal to \( K(K \geq 1) \) OFDM symbol periods, each of which consists of \( N_t = N + N_{cp} \) total sampling periods. As shown in Fig. \ref{transceiver}, $BD_i$ uses the waveform \( b[n] \) in   (\ref{wave}) to convey message \( B[n] \) in a BD symbol, for \( k = 0, ..., K - 1 \).
\begin{equation}\label{wave}
b[n] = \left\{
\begin{array}{ll}
B[n], & \text{for } n = kN_t, \ldots, \frac{2k + 1}{2} N_t - 1, \\
0, & \text{for } n = \frac{2k + 1}{2} N_t \ldots, (k + 1) N_t - 1.
\end{array} \right.
\end{equation} 

Thus, to transmit a message ‘$B[n]$’, the BD alternates its antenna impedance between two states; one state backscatters a signal with the backscatter coefficient ‘$B[n]$’ and another with no backscatter. The state transition is in the middle of each OFDM symbol period. In the figure, $B[n]=D$ to convey the random number $D$ and $B[n+1]=K_i$ to convey the identity key of $BD_i$. The waveform design aims to enable the other BD, to extract the harvested power value of the backscatter signal from the received superposed signals, as presented in the next subsection. Also, it can be easily implemented at BDs, since it is similar to the FMO waveform widely used in commercial BDs \cite{yang2017modulation}.

\subsubsection{Receiver Design}
After the backscatter of \( BD_i \), the receiver \( BD_j \) aims to remove the downlink signal and obtain the harvested power from \( BD_i \). Without loss of 
generality, let \( K = 1 \). As shown in   (\ref{5}), the downlink \( y_j^d[n] \) and backscatter \( y_j^b[n] \) signals pass through different multipath channels \( h_j \) 
and \( h_{i,j}h_i \). Only the downlink signal \( y_j^d[n] \) has a repeated CP in each 
OFDM symbol, while \( y_j^b[n] \) does not due to the backscatter waveform design (see Fig. \ref{transceiver}). Our approach uses this repeated CP to cancel \( y_j^d[n] \) and isolate the backscatter message.

To be specific, two CP parts of each OFDM symbol in the downlink signal \( y_j^d[n] \)  at \( BD_j \), are identical, i.e.,
\begin{equation}
y_j^d[n] = y_j^d[n + N], \quad n = L_{h_j} - 1, \ldots, N_{cp} - 1.
\end{equation}

In contrast, \( y_j^b[n] \) has only one CP. Thus, by subtracting:
\begin{equation}
z_j^b[n] = y_j[n] - y_j[n + N] = y_j^b[n] = h_{i,j}[n]h_i[n] B[n] \tilde{s}[n],
\end{equation}
\( BD_j \) eliminates the downlink signal and obtain the backscatter component \( h_{i,j}[n]h_i[n] B[n]\tilde{s}[n] \). Since \( BD_j \) can only measure power, it estimates the received message by evaluating the harvested power value as:
\begin{equation}
P(z_j^b[n]) = P( \eta_jh_{i,j}[n]h_i[n] B[n] \tilde{s}[n]) .
\end{equation}

\subsubsection{Synchronization}
Since the synchronization preamble in the RFS downlink signal is known to the BDs, they can use cross-correlation or methods from \cite{morelli2007synchronization} to estimate the OFDM symbol start and identify its midpoint. Even without a 
synchronization preamble, BDs can rely on the repeated CP structure \cite{yang2017modulation} or employ additional aids, such as light-based devices, for more accurate estimation \cite{xie2024integration}. Crucially, the backscatter waveform design does not require BDs to switch states precisely at the symbol midpoint. Instead, they can alternate states at any point within \([N_{cp} - 1, N + L_i - 1]\), as long as they maintain a reversed and repeated structure as in   (\ref{wave}), thus preserving the repeated CP feature. 

\begin{figure}[tp]
    % \vspace{-0.2cm}
    \centering
\includegraphics[width=\linewidth]{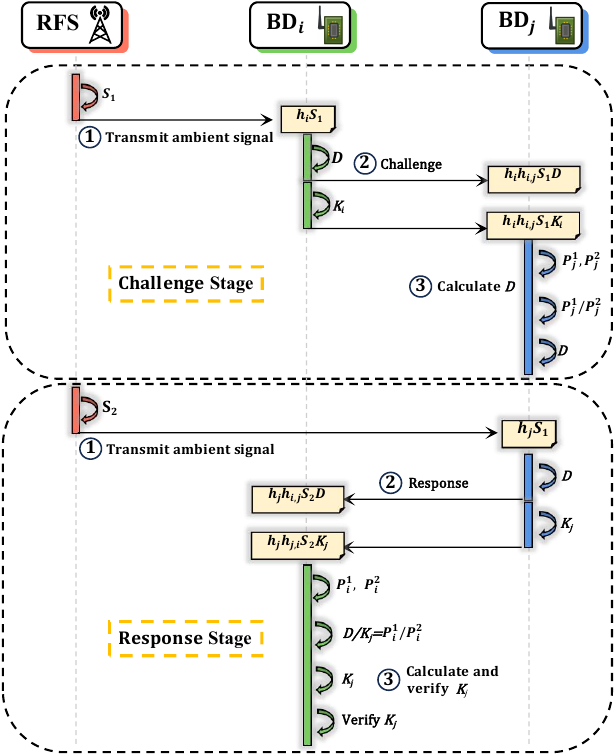}
    % \vspace{-0.1cm}
    \caption{The basic authentication procedure.}
    \label{Authentication}
\end{figure}

\subsection{One-way and Mutual Authentication}
Fig. \ref{Authentication} illustrates the basic one-way authentication procedure between $BD_i$ and $BD_j$, which consists of two stages: a) \textit{Challenge Stage} and b) \textit{Response Stage}. Without loss of generality, we assume the $BD_i$ authenticates $BD_j$ first in the authentication procedure.

\subsubsection{Challenge Stage}
In the challenge stage, $BD_i$ uses its own key $K_i$ to challenge $BD_j$ to request $K_j$ for authentication, which contains the following three steps:

\underline{\textit{Step 1-1. RFS transmits ambient signals:}} The RF source broadcasts the ambient signal $s_1(t)$, after passing the channel, $BD_i$ receives the signal as follow:
\begin{equation}
{y}_{i}(t) = h_{i}(t)  s_1(t) + w_i(t), 
\end{equation}
where $w_i(t)$ is the AWGN at $BD_i$.

\underline{\textit{Step 1-2. $BD_i$ challenges $BD_j$:}} Modulating on the incident signal from the RF source, $BD_i$ backscatters two challenge signals in two successive time, denoted as $t_1$ and $t_2$, respectively, to $BD_j$. The time interval [$t_1$,$t_2$] is limited within the channel coherence time by adjusting the length of the challenge signals in $t_1$ and $t_2$. The first signal contains a random number $D$ to inform $BD_j$ the authentication starts and secure the subsequent key exchange process. The random number can be generated using the environmental noise around the BD, which is common in existing BC systems \cite{chien2007sasi}. The second signal contains the identity key of $BD_i$ $K_i$. Note that the length of the $D$ and $K_i$ should be the same to let $t_1 = t_2$. Then, the signals arrive at $BD_j$ in $t_1$ and $t_2$ are as follows: 
%rezaei2023coding
\begin{subequations}%\label{4}
\begin{align}
{y}_{j}(t_1) &= h_{i}(t_1) h_{i,j}(t_1) D(t_1)s_1(t_1) + h_{j}(t_1) s_1(t_1) + w_j(t_1), \label{4a}\\
{y}_{i}(t_2) &= h_{i}(t_2) h_{i,j}(t_2) K_i(t_2)s_1(t_2) + h_{j}(t_2) s_1(t_2) + w_i(t_2), \label{4b}
\end{align}
\end{subequations}
where $w_j(t_1)$ and $w_j(t_2)$ is the AWGN.
%and $BD_j$ are able to separate the harvest power value of its received backscatter signal $y_j^b$ from the superposed signal ${y}_{j}$. Note that , and the method for the BD to separate the harvest power value is presented in Section \ref{Enhancement}

\underline{\textit{Step 1-3. $BD_j$ calculates $D$:}} After the above step, for easy understanding of our basic idea, let’s assume there is no noise, while the effect of noise on authentication will be examined in Section \ref{Evaluation}. Then, $BD_j$ can measure the harvest power value of the two challenge signals respectively as
\begin{subequations}%\label{5}
\begin{align}
P^1_j &= \eta_j P(h_{i}(t_1) h_{i,j}(t_1) D(t_1)s_1(t_1)), \label{7a}\\
P^2_j &= \eta_j P(h_{i}(t_2) h_{i,j}(t_2) K_i(t_2)s_1(t_2)). \label{7b}
\end{align}
\end{subequations}
%where $\eta_j$ is the energy harvesting efficiency of $B_j$.
%\textbf{Theorem 1.} \emph{

When $t_1+t_2 < T_c$, the channel fading in the harvested power of the backscatter link between two BDs is ideal \cite{shan2013phy}, i.e., $h_{i}(t_1) \approx h_{i}(t_2)$, $h_{i,j}(t_1) \approx h_v(t_2)$, and the average power of ambient signal $s(t)$ rarely changes, i.e., $P(s_1(t_1)) \approx P(s_1(t_2))$. Therefore, $BD_j$ can eliminate the effects of the channel and the signal $s(t)$ in   (\ref{7a}, \ref{7b}) by a division operation, as follows:
\begin{equation} 
P^1_j/P^2_j =  D/K_i. \label{6}
\vspace{-0.1cm}
\end{equation}
Then, $BD_j$ can estimate the random number $D$ by using its stored value of $K_i$ to multiply $\frac{D}{K_i}$.

\subsubsection{Response Stage} After the challenge, \( BD_j \) can respond to \( BD_i \) within a certain response 
time. Using the obtained random number \( D \), \( BD_j \) replies with its key \( K_j \) in the response stage, which 
involves the following steps:

\underline{\textit{Step 2-1. RFS transmits ambient signals}} The RF source broadcasts the random signal $s_2$, after passing the channel, $BD_j$ receives the signal as follow:
\begin{equation}
{y}_{j}(t) = h_{j}(t)  s_2(t) + w_j(t), 
\end{equation}
where $w_j(t)$ is the AWGN at $BD_j$.

%Note that $t_1,t_2$ and $t_3,t_4$ do not have to be in close proximity to each other, and $BD_j$ can respond within any communication time acceptable to $BD_i$.

\underline{\textit{Step 2-2. $BD_j$ responses $BD_i$ :}} Similar to \textit{Step 1-2}, $BD_j$ backscatters the estimated random number and its own key $K_j$ in two successive time slots $t_3$ and $t_4$. Refer to   (\ref{4a}-\ref{7b}), $BD_i$ can estimate the two harvest power values from the two received backscatter signals from $BD_i$ as follows:
\begin{subequations}%\label{5}
\begin{align}
P^3_i &= \eta_i P(h_{j}(t_3) h_{j,i}(t_3) D(t_3)s_2(t_3)), \label{5a}\\
P^4_i &= \eta_i P(h_{j}(t_4) h_{j,i}(t_4) K_j(t_4)s_2(t_4)). \label{5b}
\end{align}
\end{subequations}

\underline{\textit{Step 2-3. $BD_i$ calculates and verifies $BD_j$:}} Similarly, $BD_i$ can obtain the value $\frac{D}{K_i}$ using a simple division operation. Then, $BD_i$ can estimate $K_j$ using its stored value of $D$ to divide $\frac{D}{K_i}$. Due to the noise and slight differences between the channel fading in coherence time (i.e., the channel between the BDs from $t_1$ to $t_2$ and from $t_3$ to $t_4$, respectively), the calculated identity key of $BD_j$ should be similar to $K_j$ but not ideal, due to the noise and slight changes of channel fading in the coherence time . We denote the calculated identity key at $BD_i$ as $K'_j$. However, an attacker without knowledge of $D$ and $K_i$ cannot calculate $K_j$ correctly, even if it knows the authentication procedure.

Then, $BD_i$ can measure the similarity between $K_j$ and  $K'_j$ to determine the incoming authentication request is from the genuine $BD_j$ or an attacker. Given the limited computational ability of the BD, a simple and straightforward solution is to compare the Euclidean distance between $K_j$ and $K'_j$ with a predetermined threshold $\delta$:

\vspace{-0.2cm}
\begin{equation}\label{delta}
	% \label{8}
	 ||K_j-K'_j||_{l1} \underset{\mathcal{H}_{1}}{\overset{\mathcal{H}_{  0}}{\gtrless}} \delta,
\end{equation}

\noindent where the value of $\delta$ represents the threshold for making the authentication decision. $\mathcal{H}_0$ and $\mathcal{H}_1$ represent binary hypothesis tests of the authentication failure and the authentication success, respectively. Specifically, they can be denoted as:%To assess how setting different thresholds $\delta$ affects authentication performance, we examine this impact in detail in Section \ref{Evaluation}. 
\begin{equation}
\left\{
\begin{aligned}
\mathcal{H}_0: & \  K'_j =  K_a +  E_{a} \\
\mathcal{H}_1: & \  K'_j =  K_j + E_{j},
\end{aligned}
\right.
\end{equation}

\noindent where $E_{a}$ and $E_{j}$ represent estimation error at the receiver for the attacker and $BD_j$, respectively. They are caused by AWGN and the slight difference between the channel fading in coherence time. $K_a$ represents the attacker's fingerprint. %$\mathcal{H}_0$ represents that the extracted fingerprint is from an attacker and $\mathcal{H}_1$ represents that the extracted fingerprint is from the genuine $BD_j$.

After $BD_j$ passes the authentication in $BD_i$, $BD_j$ can repeat the authentication procedure to verify the authenticity of $BD_i$ for mutual authentication.

% \textit{Remark:}
% The proper selection of the threshold $\delta$ in   \ref{delta} is critical to the authentication accuracy. For example, relaxing the threshold can help the verifier correctly identify legitimate BDs, but it also increases the vulnerability of potential attackers bypassing the authentication process successfully. In Section \ref{Evaluation}, we adjust the threshold to demonstrate the trade-off between the proposed authentication scheme's ability to correctly identify legitimate BDs and the potential for mistakenly accepting unauthorized devices.

\subsection{Key Update}

In a D2D scenario, prolonged use of static device keys may lead to multiple physical-layer attacks, increasing the risk of key leakage \cite{7809147}. Meanwhile, most existing challenge-response schemes rely on a centralized server for key updates \cite{chien2007mutual}, resulting in potential single-point failures and incompatibility with AmBC systems. To address this, BDs can update their identity keys using the random number generated during each authentication. Specifically, after each successful mutual authentication, BDs broadcast the random number and their upper-layer ID (i.e., reference number). Each BD then finds the identity of the updated key according to the received reference number and updates its stored PID keys using the random number. For instance, let $D_i$ and $D_j$ be the random numbers generated by $B_i$ and $B_j$ in the authentication, respectively. Each BD can update its key through a simple geometric mean operation as follows:
\begin{subequations}%\label{5}
\begin{align}
K^{new}_i &= \sqrt{K_i \cdot D_i},\\
K^{new}_j &=\sqrt{K_j \cdot D_j}, 
\end{align}
\end{subequations}
where we can find  $\min(K_i, D_j) \leq K^{new}_i \leq \max(K_i, D_j)$ and $\min(K_j, D_j) \leq K^{new}_i \leq \max(K_j, D_j)$.  The operation of geometric averaging smoothes the value of the key and reduces the effect of extreme values in the range of [0,1] for $K_i$ and $K_j$. In addition, the key update process does not need a centralized service to allocate a new key, which enhances system scalability and reduces single points of failure.% In the next section, we will analyze the resistance of the proposed authentication procedure against the attacks defined in the adversary model \ref{Adversary Model}.
% \begin{equation}
% \left\{
% \begin{aligned}
% K^{new}_i: & \  K^{new}_i =  \sqrt{K_i \cdot D} ~~~ \min(K_i, b) \leq \sqrt{a \cdot b} \leq \max(a, b)\\
% K^{new}_j: & \  K^{new}_j = \sqrt{K_j \cdot D} ~~~ \min(a, b) \leq \sqrt{a \cdot b} \leq \max(a, b), 
% \end{aligned}
% \right.
% \end{equation}

\section{Security Analysis} \label{Security Analysis}%有位置的话可以加一张图去展示攻击者的具体的信号

The proposed authentication’s security relies on the attacker’s uncertainties about the random number, shared secret keys, and unpredictable inward channels between $BD_i$ and $BD_j$. These uncertainties prevent attackers from eavesdropping, replicating, or imitating the shared keys, ensuring resistance to impersonation, eavesdropping, replay, and counterfeiting attacks.

% In this subsection, we analyze the security of our proposed authentication procedure, focusing on its resistance against the defined attacks in the adversary model.

\subsubsection{Impersonation Attack}
The proposed authentication procedure can naturally resist impersonation attacks, as the PID keys ($K_i$, $K_j$) are unknown to them.
Impersonation attackers aim to pass the identification mechanism of the system by imitating the upper-layer identity of a legitimate BD without the ability to manipulate or imitate wireless signals. 
Therefore, the identity keys are unknown to them and PLCRA-BD can easily identify impersonation attackers by comparing whether the upper-layer identity information of the attacker claim is consistent with the shared key computed at the physical layer.

\subsubsection{Eavesdropping Attack} \label{Eavesdropping}
Eavesdropping attacks overhear communications between $BD_i$ and $BD_j$ during authentication and try to deduce \(\{{K}_{i}, K_{j}\}\). Since the challenge stage and response stage are symmetrical, we only analyze the scenario where the attacker eavesdrops on the challenge stage and tries to deduce $K_{i}$. During the signal transmission period in the challenge stage, steps 1-1 does not need to be secured since it does not reveal the shared key or CSI. In steps 1-2, the attacker can obtain:
\begin{subequations}\label{10}
\begin{align}
v^1_{i,a} &= \eta_e P(h_{i}(t_1) h_{i,e}(t_1) D(t_1)s_1(t_1)), \label{10a}\\
v^2_{i,a} &= \eta_e P(h_{i}(t_2) h_{i,e}(t_2) K_i(t_2)s_2(t_1)), \label{10b}
% v^3_{j,a} &= \eta_e P(h_{j}(t_3) h_{j,e}(t_3) D(t_3)), \label{10c}\\
% v^4_{j,a} &= \eta_e P(h_{j}(t_4) h_{j,e}(t_4) K_j(t_4)), \label{10d}
\end{align}
\end{subequations}

% Since the backscatter signal from the BD does not contain any pilot reference, the attacker cannot estimate $h_{i,e}$ at any time. 
Subsequently, We analyze the two specific eavesdropping attackers as follows.

\textbf{The naive eavesdropping attacker}: Naive eavesdropping attackers have no way to guess the shared key ${K}_{i}$ during the authentication producer because the keys exchanged between $BD_i$ and $BD_j$ are masked by the channel naturally and they do not have knowledge of ${K}_{i}$ or $D$. 
If the attacker is located at a sufficient distance \(d > \frac{\lambda}{2}\),
the attacker cannot get any information about $h_{i}$ since the channel between $h_{i}$ and $h_{a}$ is uncorrelated. The attacker also cannot estimate $h_{i,e}$ since the backscattered signal from the $BD_i$ does not contain any pilot reference. Therefore, the attacker cannot directly obtain the transmitting message from $BD_i$ by channel estimation. We further assume that the attacker knows the authentication procedure and tries to eliminate the channel fading by dividing the two factors. In this case, they can only obtain $\frac{v^1_{i,a}}{v^2_{i,a}} =  \frac{D}{K_i}$. Since the attackers do not have the knowledge of $D$ or $K_i$, they cannot solve any of them.

\textbf{The smart eavesdropping attacker}: A smart eavesdropping attacker, who is very close to $BD_i$, may get more information. If the attacker is very close to $BD_i$, we have $h_{i} \approx h_{e}$ and $h_{i,e} \approx 1$. Since the ambient signal from the source usually contains a pilot reference, the attacker can estimate $h_{e}$ and derive the approximate value of $D$ and $K_i$ from   (\ref{10a}) and (\ref{10b}) in the step 1-2, respectively. However, this kind of attack is hard to launch and easy to detect in practice. For example, if 900 MHz frequency is used, then $\frac{\lambda}{2}\approx$16.65 cm, a very short distance within which an attacker can be easily identified by legitimate users \cite{shan2013phy}.

%互信息公式可以放这里，那么这俩攻击者可能要合为一个。互信息可以推导为与距离有关的一个公式

%attempt to pass authentication in the verifier ($BD_i$ in this case) by replaying intercepted response signals to $BD_i$

%If the attacker only replays one response signal in steps 2-2, $BD_i$ cannot extract the correct $K_j$ since $K_j$ is calculated by a division operation between the two response signals. 
\subsubsection{Replay Attack}
Replay attackers are hard to succeed because the random number and the shared keys change after each authentication round. We assume that the attacker knows the authentication procedure and replays both response signals to $BD_i$ in the step 2-2, trying to pass the authentication at $BD_i$. In this case, the attacker cannot succeed since $BD_i$ accepts the request from $BD_j$ first, while the random number and the shared key change when the attacker's authentication request arrives at $BD_i$. Thus, the effectiveness of $D$ has expired, and the shared keys have been updated. As a result, $BD_i$ will calculate a wrong key using the outdated $D$ or $K_j$ received from the attacker.

\subsubsection{Counterfeiting Attack} \label{Counterfeit}
Counterfeiting attackers are hard to succeed since they cannot obtain the value of the shared key. A counterfeiting attacker mimics the authentication fingerprint of a legitimate BD, such as power and frequency, to make its RF signals indistinguishable from those of a genuine BD. However, in PLCRA-BD, the attacker cannot directly obtain the authentication fingerprint (i.e., shared keys) by eavesdropping, as analyzed before. Assuming that the attacker knows the authentication procedure in the physical layer, the attacker can only randomly guess the value of the shared keys $K_i$ and $K_j$ in steps 1-2 and 2-2, respectively. Denote attacker use the factor \(C_i\) to counterfeit $K_i$ and the factor  \(C_j\) to counterfeit $K_j$, the response signals sent by the attacker in the step 2-2 are $h_{j,e}h_{e,i} D C_i/ K_i$ and $h_{j,e}h_{e,i} C_j$. As a result, $BD_i$ will obtain $K_i/ C_i C_j$. The attacker can succeed when $K_i/ C_i C_j = K_j$, but it could need thousands of attempts, which consume significant resources. 

%优化，最小化互信息？ 

%\input{s4_secan}
%信道的非理想特性

\section{Performance Evaluation} \label{Evaluation} 
This section presents the simulation setup and evaluates the proposed authentication scheme’s performance in terms of effectiveness, accuracy, robustness, and efficiency.

% by addressing the following research questions. 
% \begin{itemize}
% \item \textbf{Authentication Effectiveness:} Can our proposed authentication scheme successfully distinguish legitimate devices and illegal devices? Can our proposed authentication scheme and baseline be used in AmBC systems?

% \item \textbf{Authenticaion Accuracy:} How accurate is our proposed authenticaion scheme under different parameter setting and channel conditions?%, inculding SNR, environment, velocity, and distance between devices
% \item \textbf{Attack Robustness:} How robust is AuthScatter when faced with various wireless attacks?

% \item \textbf{Efficiency:} %What is the power consumption and latency of our proposed authentication scheme? 
% How efficient is our proposed authentication scheme when compared with traditional wireless attacks in terms of latency and power consumption? 
% \end{itemize}

% \begin{figure}[tp]
%     \centering
%     \includegraphics[width=0.7\linewidth]{figs/simulation_setting.pdf}

%     \caption{Simulation setting.}
%     \label{Simulation setting}

% \end{figure}

\vspace{-0.3cm}
\subsection{Simulation Setup}
\subsubsection{Simulation Settings} 

Monte Carlo simulations are conducted in a Matlab platform to analyze the authentication performance under different system parameters, scenarios, and attacks. We consider an AmBC system composed of a fixed ambient RF source (RFS) and multiple BDs, where some BDs are stationary and others may be mobile. In the simulation, two paired BDs exploit the ambient signal from the RFS to perform authentication, including one-way authentication and mutual authentication. 
Additionally, we introduce an attacker positioned near a legitimate BD, aiming to eavesdrop on the communication link or circumvent the authentication process between the two BDs. Following the adversary model described in Subsection \ref{Adversary Model}, this attacker may engage in impersonation, eavesdropping, replay, or counterfeiting attacks. The default parameter settings used throughout the simulations, as presented in Table \ref{tab:tab2_notations}, referred to those commonly adopted in authentication schemes and AmBC systems \cite{shan2013phy,yang2017modulation,BCAuth,li2022security}.%,liu2013ambient,wang2021physical

% By default, the carrier signal frequency and the maximum average power of the RFS signal is set at 900 MHz and \(P_{\rm T}=1\) dBm, respectively. The received noise power at a device is set to be $\sigma^2 = -30 \, \mathrm{dBm}$. The distance between the RF source and two BDs $B_i$, $B_j$
% is set as $5m $ and the distance between the two BDs is set as $3m$, respectively. The channel is modeled by Rayleigh fading, and its the 
% channel gains are set as $10^{-2}d^{-2}$. The default setting comes from the typical authentication scenario of the BC system referring to previous works \cite{BCAuth} and \cite{wang2021physical}.
% To ensure reliability, 1,000 repeated authentication processes are performed for each simulation. 

\begin{table}
    \caption{Parameter Settings}
    \centering
    \begin{tabular}{lll}
        \toprule
        \textbf{ Notation } & \textbf{Description}& \textbf{Setting} \\ 
        \midrule
 
         \(f_{\rm c}\) & carrier signal frequency  & 900 Mhz \\  
         $\sigma^2$ & received noise power  &-30 dBm\\  
         \(P_{\rm T}\) & maximum average power of the RFS  & 1 dBm \\ 
         \(D_{\rm i}\),\(D_{\rm j}\) & the distance between RFS and $B_i$,$B_j$  & 3 m \\ 
         \(D_{\rm i,j}\) & the distance between BDs  & [1,10] m \\ 
        \(D_{\rm a}\) & the distance between a BD and a attacker  & [0.1,2] m \\ 
         \(v_{\rm i,j}\) & the relative spend between BDs  & [0,30] m/s \\ 
        keylength& the length of the shared PID key in BDs  & [5,30] \\ 
         \(h\) & channel model  & Rayleigh fading \\ 
         \(|h|\) & channel gain  & $10^{-2}d^{-2}$ \\ 
         \(N_{auth}\) & authentication number  & 1000 \\ 
                 
         \bottomrule
    
    \end{tabular}
    
    \label{tab:tab2_notations}

\end{table}

\subsubsection{Baselines}
Two conventional authentication schemes, \textbf{Baseline1} \cite{chien2007sasi} and \textbf{Baseline2} \cite{wang2012server}, are compared with PLCRA-BD.
\textbf{Baseline1} uses a lightweight handshake protocol that shares a secret key between devices through a combination of random numbers and XOR operations. This scheme imposes minimal computational and hardware requirements, making it suitable for resource-constrained BDs, but it provides only limited security protection. In contrast, \textbf{Baseline2} utilizes hash functions to secure the key exchange process, delivering stronger security but increasing complexity. As a result, it is generally impractical for severely constrained devices. We do not compare with existing PLA schemes because they cannot be implemented between BDs in an AmBC system.

\subsubsection{Evaluation Metrics} 

Following previous PLA works in \cite{BCAuth,shan2013phy, yang2024batchauth}, we evaluate authentication accuracy under spoofing attacks using: \textbf{1) True positive rate (TPR)}, defined as the rate that a genuine BD is truly accepted, and \textbf{2) False positive rate (FPR)}, defined as the rate at which attackers are incorrectly accepted. Then, the \textbf{receiver operating characteristic (ROC)} curve is used to illustrate the trade-off between TPR and FPR across different thresholds $\delta$ in   (\ref{delta}), where its range is determined through empirical training. In addition, the mutual information between signals received from legitimate devices and attackers is measured to reflect the secrecy of authentication schemes, referred to as  \textbf{leaked information (LI)} \cite{li2022security}. Regarding efficiency, the  \textbf{authentication latency} and \textbf{power consumption} are considered. 

\begin{figure}[t]
    \centering
    \includegraphics[width=0.7\linewidth]{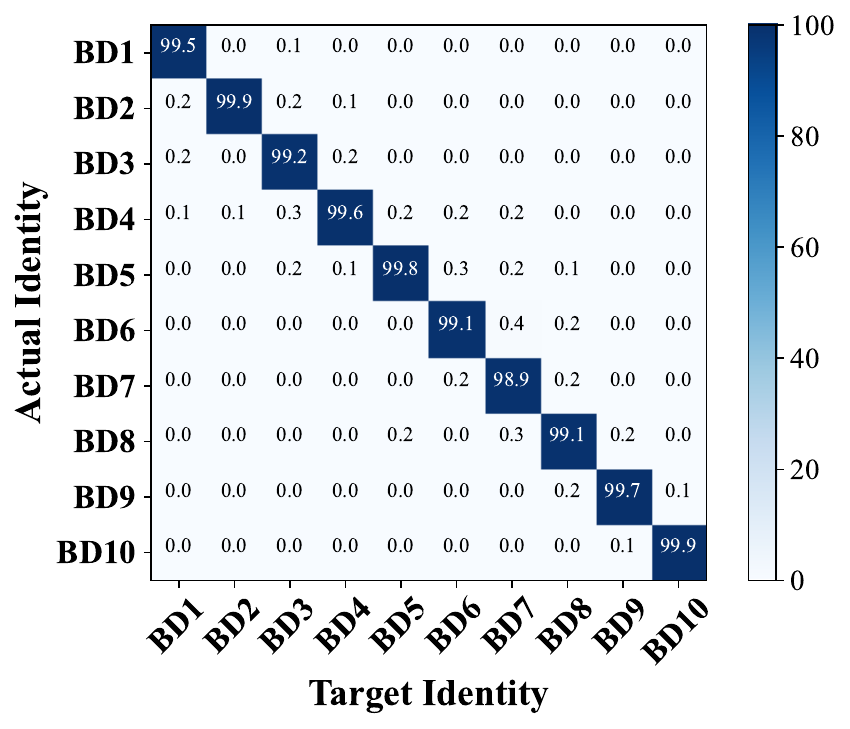}
    \caption{Confusion matrix for BD authentication with each box showing the percentage of devices identifying as the corresponding identity.}
    %\caption{Confusion matrix in identifying legitimate BDs, with each box showing the percentage of devices classified as the corresponding class.}
    \label{Eff}
    %\vspace{-0.6cm}
\end{figure}

\subsection{Effectiveness and complexity analysis}
%\subsubsection{Effectiveness of our proposed authentication scheme} 
\subsubsection{Effectiveness analysis}
We conducted a basic authentication experiment to evaluate the effectiveness of PLCRA-BD in identifying legitimate BDs. The experiment follows a straightforward principle: a prover BD is identified as the identity (i.e., actual device) whose PID key most closely matches its own. As illustrated in Fig.~\ref{Eff}, we tested 10 legitimate BDs, each assigned a PID key of length 10. The results show that the authentication accuracy for all legitimate BDs exceeds 99\%, demonstrating the scheme's ability to accurately distinguish between devices. Although a small probability of false identification exists for devices with similar PID keys, this issue can be mitigated by increasing the key length, thereby enhancing key differentiation and improving device identification.

% \textbf{Result}:
% \textbf{Analysis}:

\begin{table}[t]
\centering
\caption{Complexity comparison of PLCRA-BD and baselines.}
\label{dataset}
\resizebox{\columnwidth}{!}{%
\begin{tabular}{c|c|c|c|c}
\toprule[1.5pt]
Scheme & \makecell{Time \\ Complexity} & \makecell{ Decoding\\ Avoidance} & \makecell{Physical Layer \\ Implementation} & \makecell{AmBC \\ Support} \\ \midrule
Baseline1 & $O(N)$ & \emptycirc & \fullcirc & \fullcirc \\
Baseline2      & $O(N)$ & \emptycirc & \emptycirc & \emptycirc \\ \midrule
Ours      & $O(1)$ &  \fullcirc &  \fullcirc &  \fullcirc \\ 
\bottomrule[1.5pt]
\end{tabular} \label{EFF_table}% 
}
% \vspace{-2mm}
\end{table}
\subsubsection{Complexity analysis} \label{Complexity analysis} 
Table~\ref{EFF_table} compares the computational complexity and device requirements of PLCRA-BD against two baseline schemes. Both baselines incur $O(n)$ complexity due to XOR and hash-based encryption operations. In contrast, our scheme requires transmitting only constant-length bits for the random number and the PID key, achieving $O(1)$ complexity.
Additionally, Baseline1 and Baseline2 must decode the transmitted signal to extract the random number and key, thereby increasing power consumption. Our scheme avoids this step entirely by deriving fingerprints directly from the energy harvester’s output, eliminating the need for decoding.
Furthermore, the baseline2 relies on hash operations typically implemented at the upper software layer, such as for password verification or digital signatures, which is impractical in AmBC systems given the limited capabilities of BDs. In comparison, both the simple XOR operation in Baseline1 and the backscatter operations in our proposed scheme function at the physical layer, aligning with passive BDs.

\subsection{Authenticaion Accuracy} \label{Accuracy}
This subsection simulates an impersonation attack where an attacker uses intercepted upper-layer IDs to bypass authentication. We evaluate the accuracy of PLCRA-BD under varying conditions, including key length, SNR, BD distance, and velocity.

\begin{figure}[!tb]
    \centering
    %\vspace{-0.2cm}
    \subfigure[One way Authentication]{
    \includegraphics[width=0.47\columnwidth]{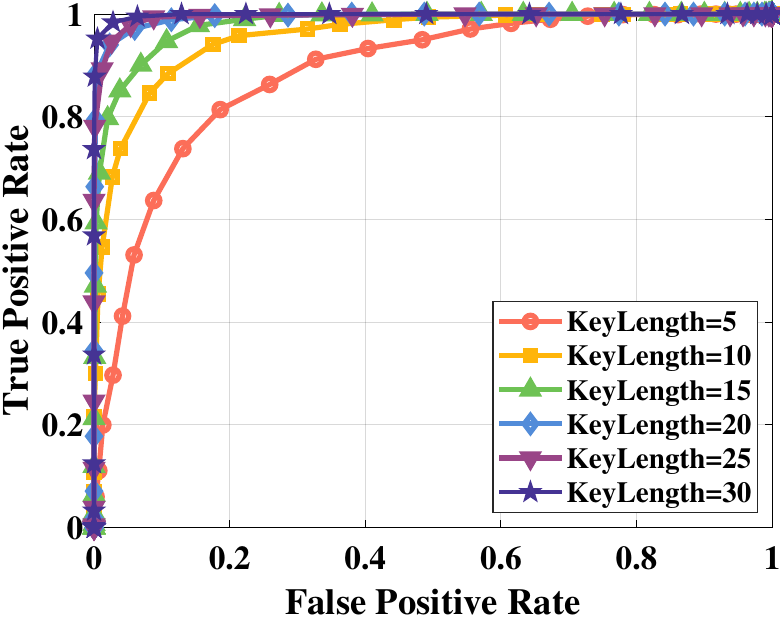}
    \label{Onewaykeylength}
        }
    \hspace{-6 mm}
    \subfigure[Mutual Authentication]{
        \includegraphics[width=0.47\columnwidth]{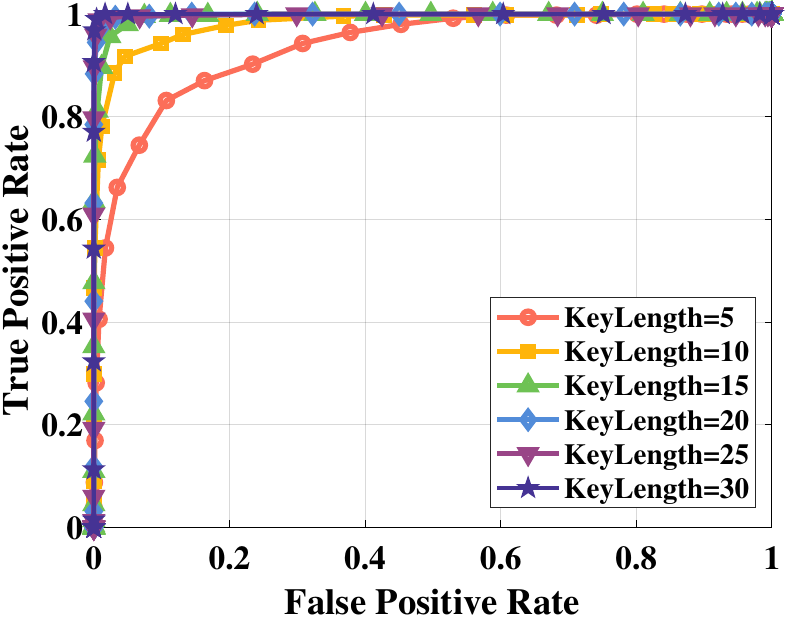}
        \label{fig:keymutual}
    }
    % \vspace{-2mm}
    \caption{ROC of (a) one-way authentication and (b) mutual authentication, with various keylength.}
    \label{key}
    % \vspace{-0.6cm}
\end{figure}

\subsubsection{Impact of key length} 

% \textbf{Result}:
% \textbf{Analysis}:
Fig.~\ref{key} presents the ROC curves for one-way and mutual authentication under varying lengths of the shared keys $K_i$ and $K_j$. A larger area under the ROC curve corresponds to higher authentication accuracy. As the key length increases, the ROC curves consistently shift upward, indicating improved performance. This enhancement arises because longer keys better differentiate legitimate BDs from attackers, making it increasingly difficult for adversaries to accurately guess the keys.
Furthermore, comparing the ROC curves in Fig.~\ref{Onewaykeylength} and Fig.~\ref{fig:keymutual} shows that the accuracy of mutual authentication outperforms that of one-way authentication. This is due to the stricter requirement in mutual authentication, where attackers must correctly guess both $K_i$ and $K_j$ simultaneously, a significantly more challenging task than guessing a single key.

\begin{figure}[!tb]
    \centering
    \vspace{-0.4cm}
    \subfigure[One-way Authentication]{
        \includegraphics[width=0.47\columnwidth]{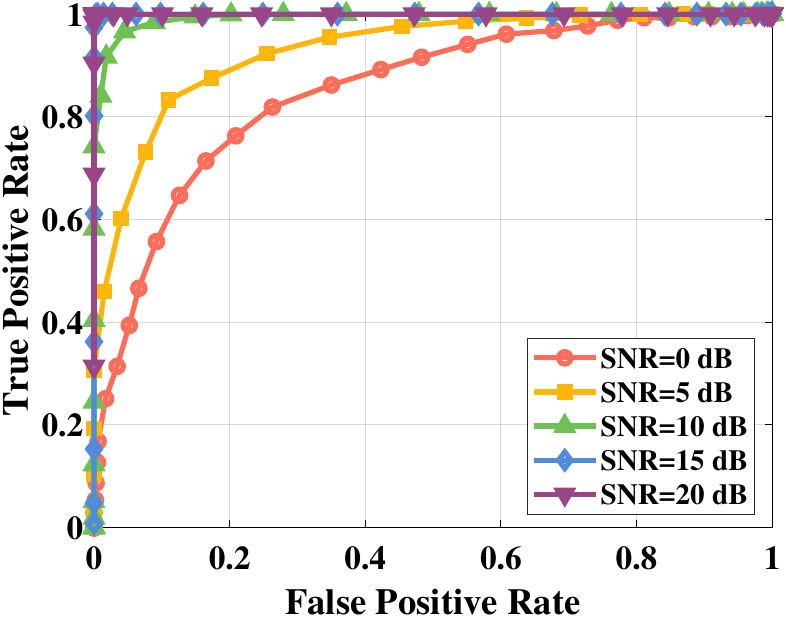}
        \label{}
    }
    \hspace{-6mm}
    \subfigure[Mutual Authentication]{
        \includegraphics[width=0.47\columnwidth]{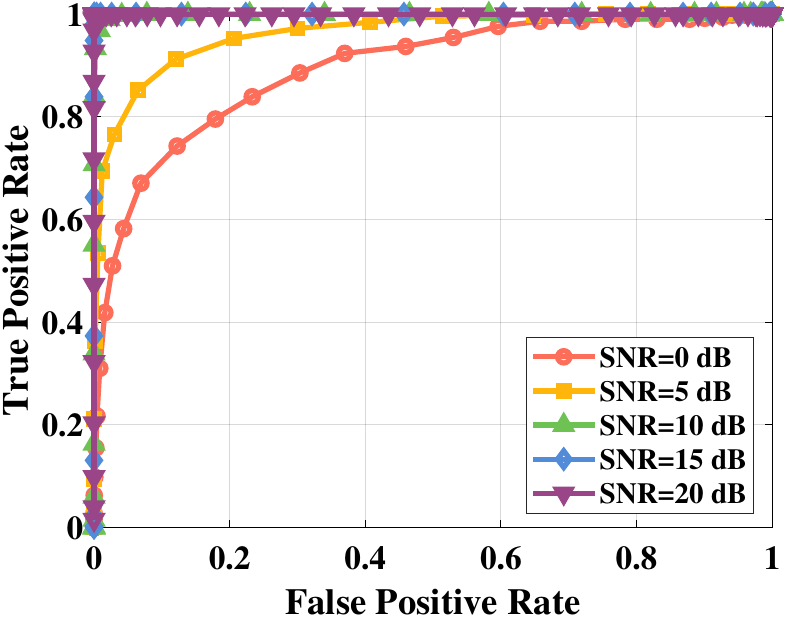}
        \label{}
    }
    %\vspace{-2mm}
    \caption{ROC of (a) one-way authentication and (b) mutual authentication, under various SNR values.}
    \label{SNR}
    % \vspace{-1mm}
\end{figure}

\subsubsection{Impact of noise}

Fig.~\ref{SNR} illustrates the ROC performance at a receiver BD for SNR values ranging from 0 to 20\,dB, with the SNR adjusted by varying the transmitting power level. The results indicate that authentication accuracy improves as SNR increases, since higher SNR facilitates more accurate key extraction and identification.
Notably, the TPR remains at 1 regardless of changes in the FPR when the SNR exceeds 15\,dB for one-way authentication and 10\,dB for mutual authentication. This result confirms that our scheme can achieve high authentication accuracy by increasing SNR, a goal attainable through practical measures such as reducing the distance between BDs or increasing the transmitting power of the RFS.

\subsubsection{Impact of the distance between BDs}

\begin{figure}[!tb]
    \centering
    %\vspace{-0.1cm}
    \subfigure[One-way Authentication]{
        \includegraphics[width=0.47\columnwidth]{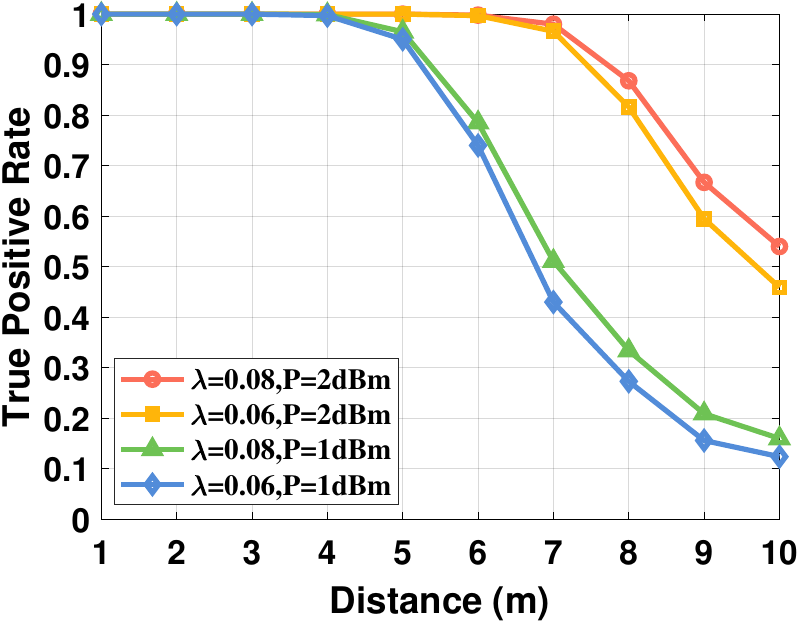}
        \label{}
    }
    \hspace{-6mm}
    \subfigure[Mutual Authentication]{
        \includegraphics[width=0.47\columnwidth]{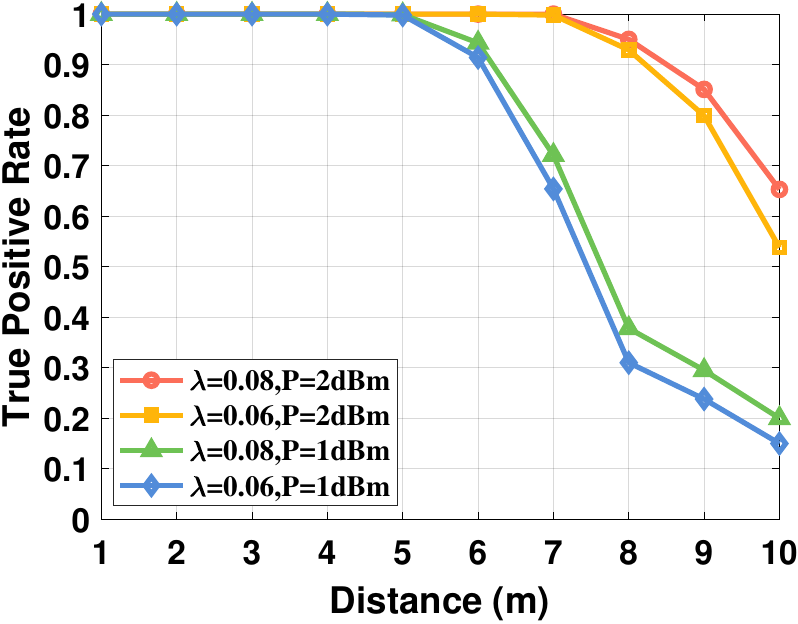}
        \label{}
    }
    %\vspace{-2mm}
    \caption{TPR vs. the distance between BDs of (a) one-way authentication and (b) mutual authentication.}
    \label{D}
    % \vspace{-1mm}
\end{figure}

Fig.~\ref{D} illustrates the effect of the distance between BDs on authentication accuracy. The TPR decreases as the distance increases, primarily due to greater large-scale attenuation in the inward channel between the BDs, which lowers the SNR at the verifier BD.
The results also indicate that TPR can be improved by increasing the RFS transmitting power or by lowering the FPR limit. Increasing the transmitting power is a standard approach in wireless systems, while lowering the FPR threshold involves a trade-off, as it reduces the scheme’s ability to correctly identify attackers.
For subsequent simulations, the distance between the two BDs is fixed at 3\,m.

%\cite{juels2006rfid}

% In addition, the high TPR values (above 0.9 for RFS and 0.6 for BD authentication) at distances up to 3 meters indicate AuthScatter's good deployment potential.

\begin{figure}[!tb]
    \centering
    % \vspace{-0.1cm}
    \subfigure[BD mobility]{
        \includegraphics[width=0.47\columnwidth]{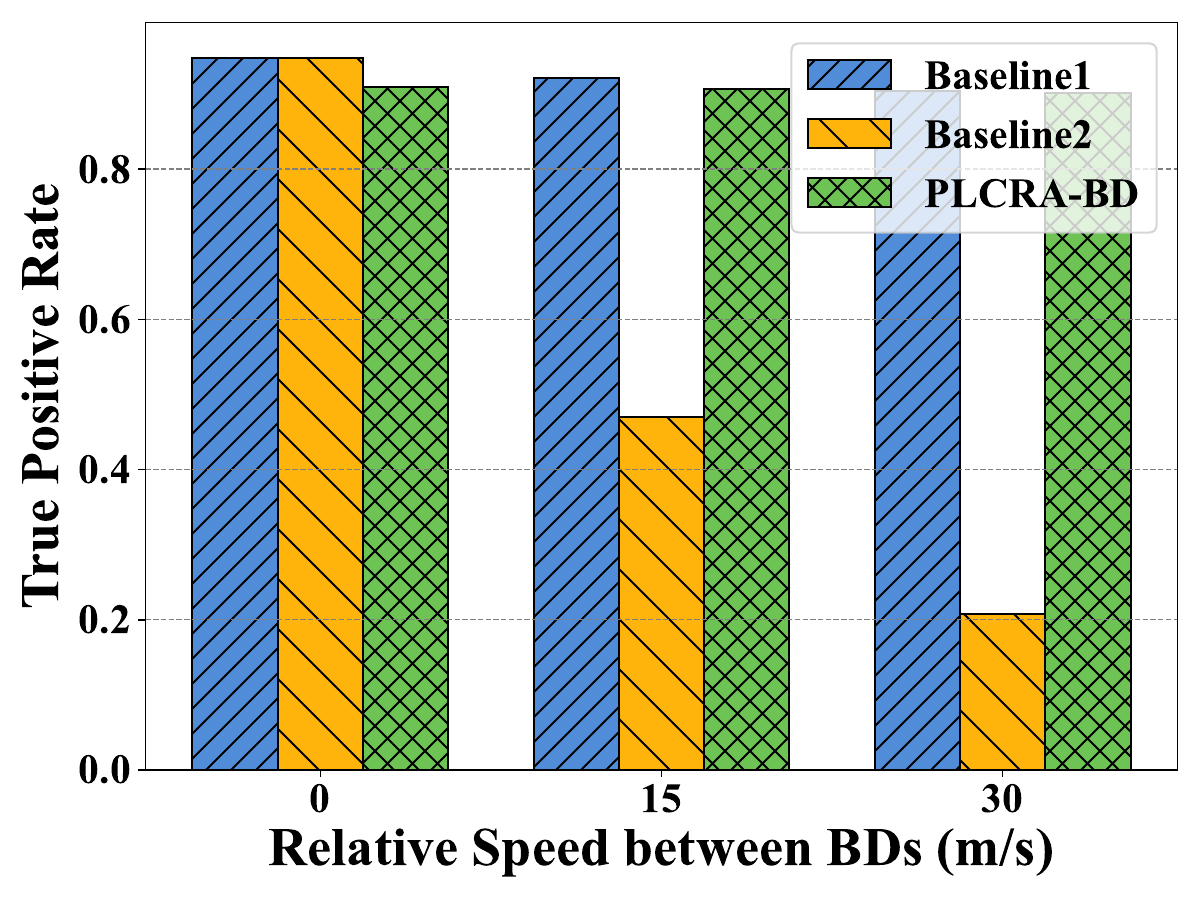}
        \label{BD mobility}
    }
    \hspace{-6mm}
    \subfigure[Environment Mobility]{
        \includegraphics[width=0.47\columnwidth]{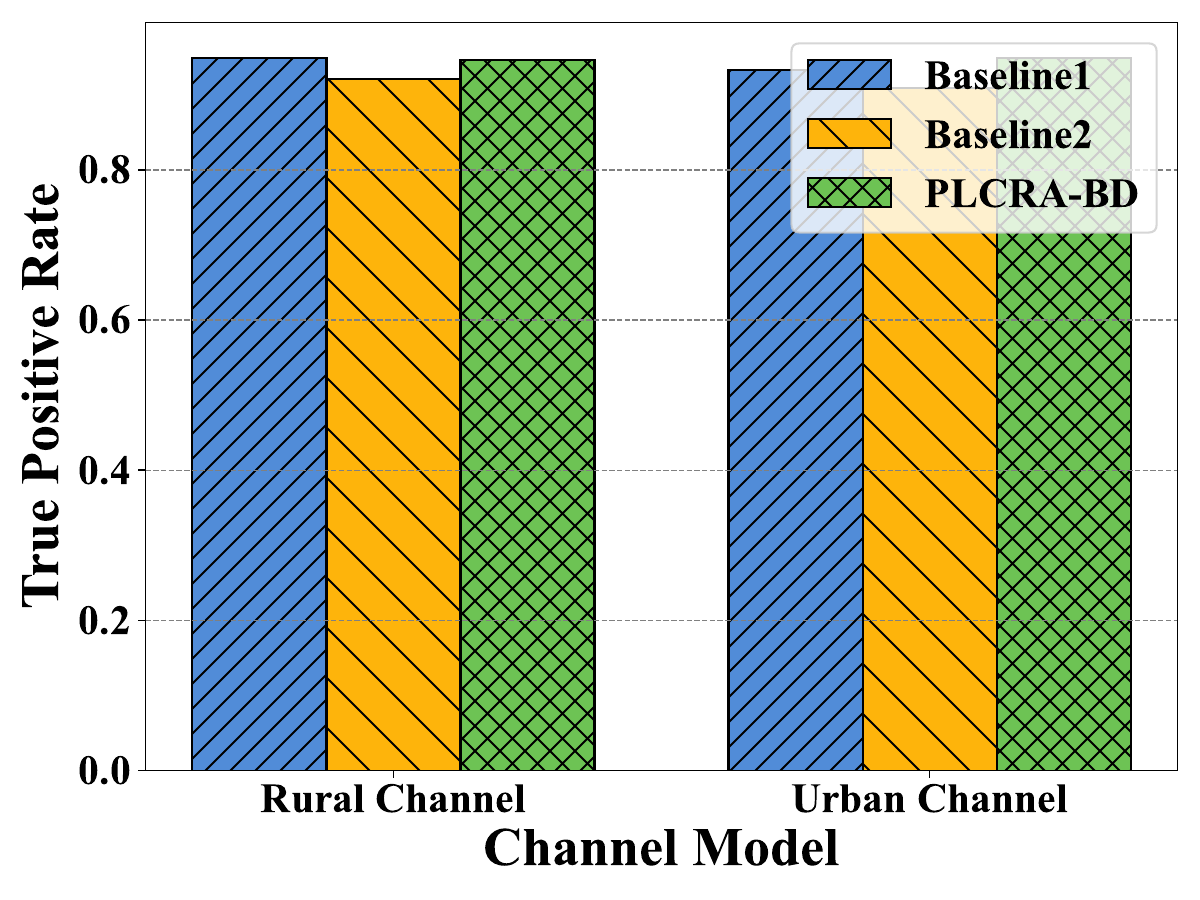}
        \label{environment mobility.}
    }
    \caption{TPR comparison between PLCRA-BD and baselines with (a) BD mobility and (b) environment mobility. The TPR is tested at a FPR limit of 0.}
    \label{velocity}
\end{figure}

\subsubsection{Impact of the BD and environment mobility} 
Fig.~\ref{BD mobility} presents the effect of BD speed on the TPR. It can be seen that PLCRA-BD maintains a nearly constant TPR even at speeds up to 30~m/s, indicating robust support for BD mobility. Meanwhile, Fig.~\ref{environment mobility.} shows the impact of different channel conditions, where we employ the rural/urban channel models defined in \cite{3gpp-ts-45005-v790}, characterized by 4/12 multipaths, fixed amplitudes, and random phase shifts. The TPR of PLCRA-BD remains stable under these diverse environmental conditions, demonstrating adaptability to environmental mobility. Furthermore, both subfigures reveal that the TPR values of Baseline1 and PLCRA-BD are closely aligned, while those of Baseline2 are lower. This confirms that PLCRA-BD achieves authentication accuracy on par with commonly employed traditional methods \cite{chien2007mutual,wang2012server,cho2015consideration,gao2019secure,chien2007sasi,1599066,sun2009security}.

\vspace{-4mm}
\subsection{Attack Robustness} \label{Robustness}
This subsection evaluates the robustness of the authentication procedure against wireless attacks, including eavesdropping, replay, and counterfeiting, and compares it with the baseline methods.

\begin{figure}[!tb]
    \centering
    \vspace{-0.1cm}
    \subfigure[Different eavesdropping attacks]{
        \includegraphics[width=0.47\columnwidth]{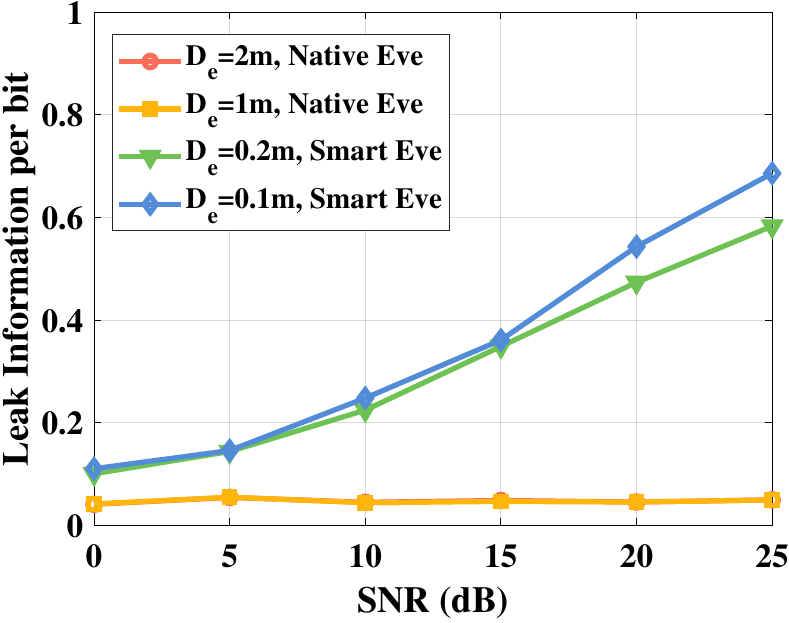}
        \label{DiffEav}
    }
    \hspace{-6mm}
    \subfigure[Different schemes]{
        \includegraphics[width=0.47\columnwidth]{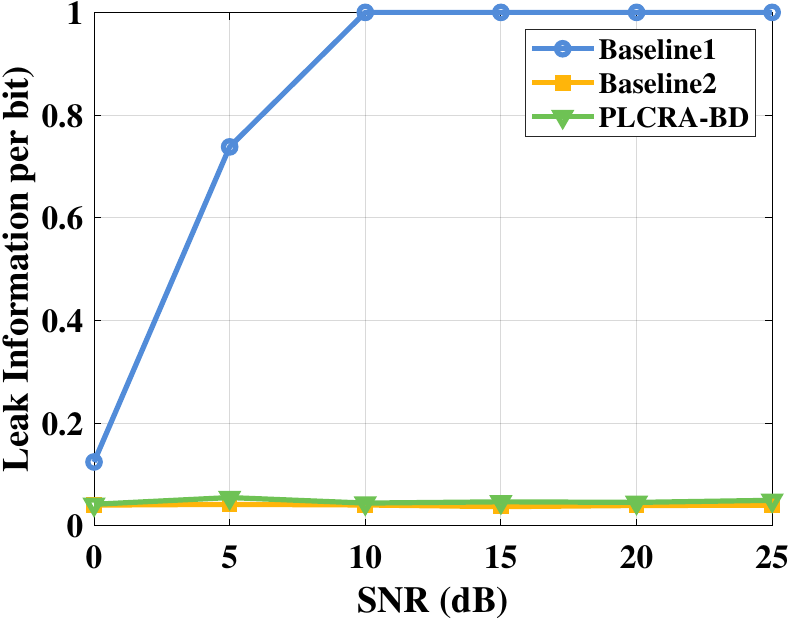}
        \label{DiffSch}
    }
    %\vspace{-2mm}
    \caption{Leaked information vs. SNR under eavesdropping attacks}
    \label{Eav}
    \vspace{-2mm}
\end{figure}

\subsubsection{Robustness against eavesdropping attacks}

% \textbf{Result}:
% \textbf{Analysis}:
Fig. \ref{DiffEav} illustrates the variation of LI with SNR values for PLCRA-BD under different eavesdropping attacks. In naive eavesdropping scenarios, the LI is minimal at approximately 0.04 and remains nearly constant despite increasing SNR or decreasing distance $D_a$. This demonstrates the strong robustness of our scheme against naive eavesdropping attacks. Conversely, the figure shows a higher LI (above 0.6) in smart eavesdropping cases as the SNR increases or $D_a$ decreases, indicating the potential vulnerability of the scheme to smart eavesdropping. However, smart eavesdroppers are uncommon in practice and are also challenging to address with existing PLA methods in BC systems.

Fig. \ref{DiffSch} compares the LI of PLCRA-BD with the baselines under naive eavesdropping attacks. The results show that the LI values of Baseline2 are very low. This is because the high secrecy of hash encryption in Baseline2 prevents attackers from obtaining the genuine message during authentication, leaving them to rely solely on random guesses of the shared keys. Meanwhile, the close LI values of Baseline2 and PLCRA-BD highlight the strong secrecy of our scheme. In contrast, Baseline1 is unable to resist naive eavesdropping attacks, as an eavesdropper can easily break the XOR encryption by comparing the exchanged ciphertext with the transmitted plaintext, thereby recovering the secret key.

\subsubsection{Robustness against replay attacks}
\begin{figure}[!tb]
    \centering
    % \vspace{-0.1cm}
    \subfigure[One-way Authentication]{
        \includegraphics[width=0.48\columnwidth]{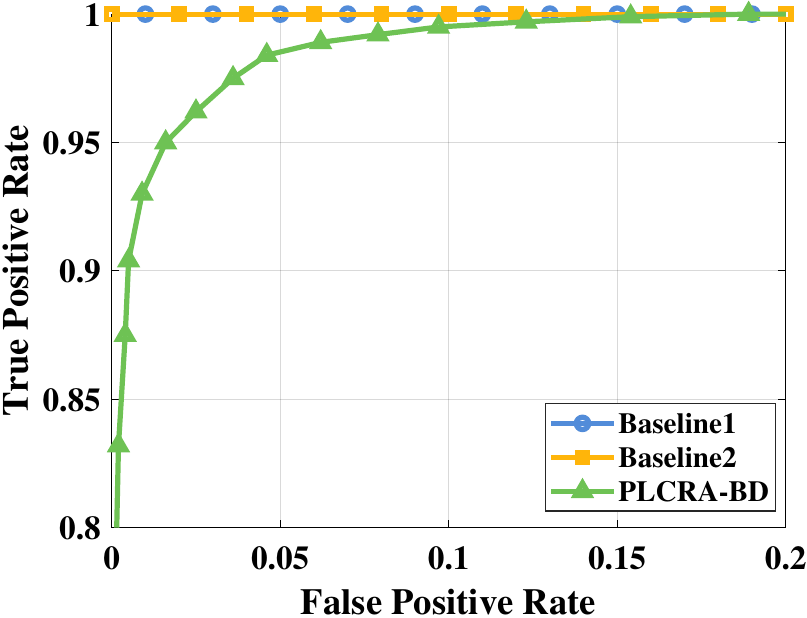}
        \label{}
    }
    \hspace{-6mm}
    \subfigure[Mutual Authentication]{
        \includegraphics[width=0.48\columnwidth]{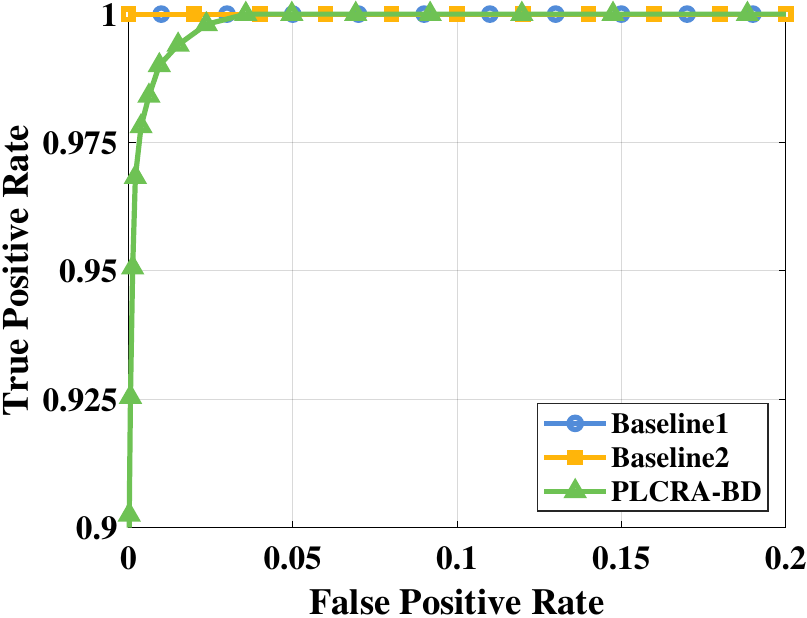}
        \label{}
    }
    %\vspace{-2mm}
    \caption{ROC of (a) one-way authentication and (b) mutual authentication, under replay attacks.}
    \label{replay}
    % \vspace{-1mm}
\end{figure}

% \textbf{Result}:
% \textbf{Analysis}:
Fig.~\ref{replay} compares the ROC performance of PLCRA-BD with baseline schemes under replay attacks. The results show that PLCRA-BD achieves a TPR of approximately 0.95 for one-way authentication and 0.99 for mutual authentication at a 0.02 FPR, indicating that our scheme can effectively prevent most replay attackers. In contrast, the baselines exhibit even greater robustness against replay attacks. This is primarily because these schemes are less susceptible to environmental factors, whereas the exchanged keys in PLCRA-BD are more vulnerable to channel fading and noise.

% \begin{figure}[!tb]
%     \centering
%     % \vspace{-0.1cm}
%     \subfigure[One-way Authentication]{
%         \includegraphics[width=0.45\columnwidth]{figs/onewaycounterfeit.pdf}
% }
%         \label{}
%     }
%     \hspace{-6mm}
%     \subfigure[Mutual Authentication]{
%         \includegraphics[width=0.45\columnwidth]{figs/mutualcounterfeit.pdf}
%         \label{}
%     }
%     %\vspace{-2mm}
%     \caption{ROC of (a) one-way authentication and (b) mutual authentication, under counterfeiting attacks.}
%     \label{Counterfeiting}
%     % \vspace{-1mm}
% \end{figure}

\begin{figure}[!tb]
    \centering
    % \vspace{-0.1cm}
    \subfigure[One-way Authentication]{
        \includegraphics[width=0.455\columnwidth]{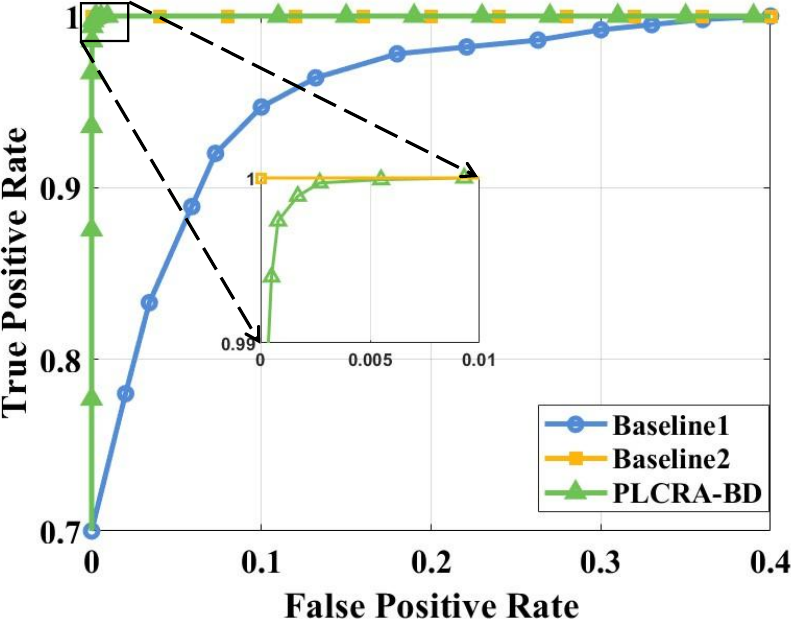}
        \label{}
    }
    \hspace{-6mm}
    \subfigure[Mutual Authentication]{
        \includegraphics[width=0.455\columnwidth]{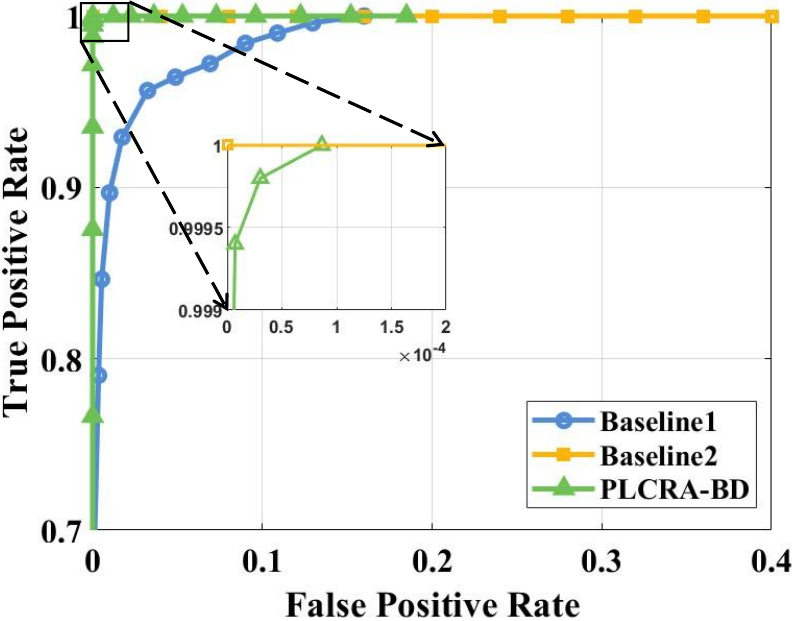}
        \label{}
    }
    \caption{ROC of (a) one-way authentication and (b) mutual authentication, under counterfeiting attacks.}
    \vspace{-2mm}
    \label{Counterfeiting}

\end{figure}
\subsubsection{Robustness against counterfeiting attacks}
Fig. \ref{Counterfeiting} compares the ROC performance of PLCRA-BD with baselines under counterfeiting attacks. Of the three schemes, Baseline1 is the least robust to counterfeiting attacks, due to the fact that the key of Baseline1 can be easily estimated by eavesdropping attacks. Using estimated keys, counterfeiting attackers can increase their FPR performance. In contrast, the counterfeiting attackers in PLCRA-BD have approximately  0.005 and $10^{-4}$ FPR with 1 TPR limit for one-way and mutual authentication, respectively. This extremely low FPR demonstrates the strong robustness of our scheme against counterfeiting attacks. Although Baseline2 achieves the best performance, it cannot be achieved in AmBC systems due to its high complexity requirements. As a result, our scheme has close performance with the Baseline2 and can be implemented in AmBC systems.

\begin{figure}[!tb]
    \centering
    \vspace{-0.1cm}
    \subfigure[Authentication Latency]{
        \includegraphics[width=0.47\columnwidth]{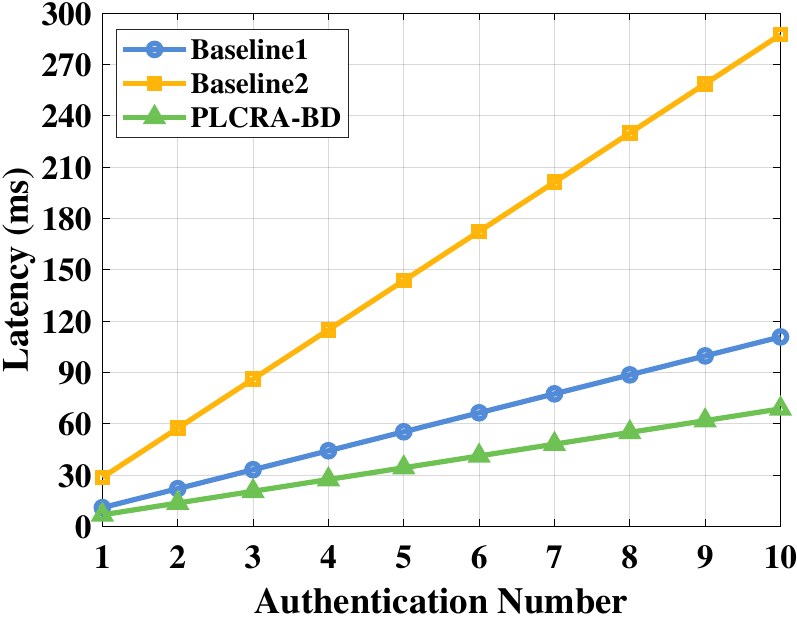}
        \label{Latency}
    }
    \hspace{-6mm}
    \subfigure[Power Consumption]{
        \includegraphics[width=0.47\columnwidth]{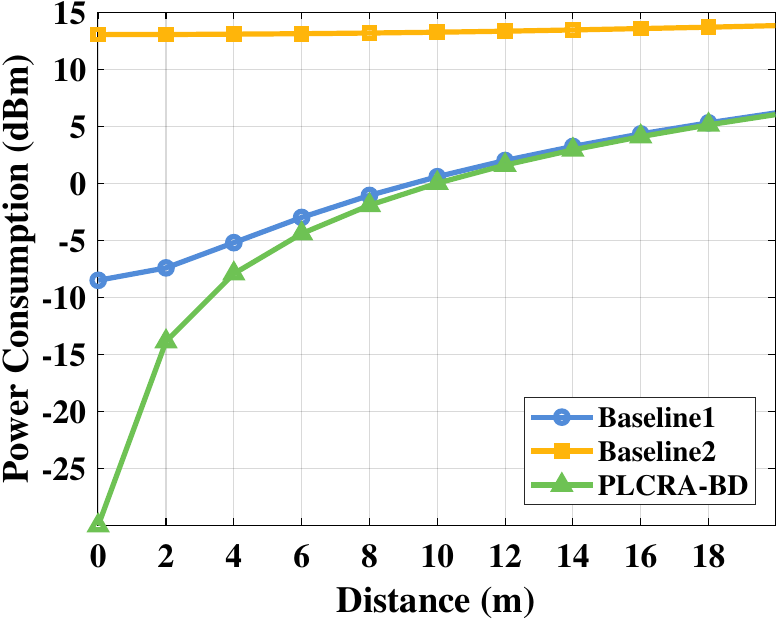}
        \label{Power}
    }
    %\vspace{-2mm}
    \caption{Efficiency comparison between PLCRA-BD and baselines.}
    \label{SNR}
    \vspace{-2mm}
\end{figure}

\vspace{-3mm}
\subsection{Efficiency}
The efficiency is evaluated under a one-way authentication procedure since the efficiency of mutual authentication is a simple linear relationship with the one-way authentication.

\subsubsection{Authentication latency}
Fig. \ref{Latency} compares the authentication latency of PLCRA-BD with the baselines for different authentication numbers. For PLCRA-BD, we consider signal transmission time $T_{\text{tx}}$, random number generation time $T_{\text{rand}}$, and the key verifying time $T_{\text{verify}}$. $T_{\text{Our}} = 4T_{\text{tx}} + T_{\text{rand}} + T_{\text{verify }}$. For baseline1, we consider the signal transmission time, the random number generation time, the verifying time, bit operation time $T_{\text{XOR}}$, and signal decoding time $T_{\text{decoding}}$. $T_{\text{baseline1}} = 4T_{\text{tx}} + T_{\text{rand}} + T_{\text{verify}}  +2T_{\text{XOR}}+4T_{\text{decoding}}$. For baseline2, we consider the signal generation time $T_{\text{gen}}$, hash encryption time $T_{\text{hash}}$, the signal transmission time, the random number generation time, and signal decoding time $T_{\text{Baseline2}} = 2T_{\text{gen}} + 2T_{\text{tx}} + 2T_{\text{hash}}+T_{\text{rand}} + T_{\text{verify}} +2T_{\text{decoding}}$. From Fig. \ref{Latency}, it can be found that PLCRA-BD achieves the lowest authentication latency.% Compared to baseline1, PLCRA-BD can save time for XOR and signal decoding operations, and this could be important for large-scale BD scenarios.

\subsubsection{Power consumption}
Fig. \ref{Power} provides an estimation of power consumption for PLCRA-BD and the baselines relative to the distance between BDs. The analysis includes computation, baseband, and RF transmission power. For a low-power BD, a simple decoding operation consumes approximately 0.03 mW, and an XOR operation consumes 0.01 mW, whereas a regular device requires 5–10 mW for hash encryption. The RF power at the transmitter BD must meet the SNR requirement of 10 dB for all schemes. As shown in Fig. \ref{Power}, PLCRA-BD consumes less power than baseline1 and baseline2, demonstrating its superiority for practical deployment.
\vspace{-3mm}
\section{Conclusion}\label{Conclusion}

In this paper, we introduced PLCRA-BD, a novel physical layer authentication scheme between BDs in AmBC systems based on challenge and response, achieving high-mobility support and robust security. First, we design a fingerprint embedding method and an integrated transceiver mode in resource-constraint BDs, which enables BDs to transmit and extract shared PID-bearing signals in a low-complexity way without the interference of the ambient signals.
Building on this, a challenge-response authentication procedure is proposed that enables BDs to exchange shared PID keys in a secure way by exploiting a random number and channel fading, achieving both one-way and mutual authentication. The key update design further mitigates the risk of key leakage. Then, we theoretically analyze the resistance of the authentication procedure against impersonation, eavesdropping, replay, and counterfeiting attacks. Finally, the performance of PLCRA-BD was comprehensively evaluated under various system settings using numerical simulations. It demonstrates that the scheme can be easily implemented in resource-constraint AmBC systems with desirable accuracy, advanced robustness, and superior efficiency compared with conventional authentication schemes.% In the future, continuing research to prototype the authentication protocol in existing commercial BDs need further investigation. In addition, with the intensive deployment of BDs in AmBC systems, lightweight group authentication between multiple BDs is worth investigating.

%从模式的角度去叙述，而不是秘钥，秘钥是导致了存储的反射模式，这个反射模式隐藏在混乱的信道之中

%系统图，符号表，算法table*2，帧设计图，接收者设计图

%固定的相位噪声可以由一个移位展示出来
%\section*{Acknowledgment}
%This work was supported in part by the National Natural Science Foundation of China under Grant 62072351; in part by the Key Research Project of Shaanxi Natural Science Foundation under Grant 2023-JC-ZD-35; in part by the open research project of ZheJiang Lab under grant 2021PD0AB01; in part by the Academy of Finland under Grant 345072 and Grant 350464; and in part by the U.S. National Science of Foundation through the Networking Technology and Systems (NeTS) Program under Award 2131507. (Corresponding author: Zheng Yan.)

% \newpage
% \bibliographystyle{IEEEtran}
% \bibliography{IEEEabrv,reference}

\bibliographystyle{plain}
\bibliography{reference}

\end{document}